\documentclass{article}

\usepackage{arxiv}

\usepackage[utf8]{inputenc} 
\usepackage[T1]{fontenc}    
\usepackage{hyperref}       
\usepackage{url}            
\usepackage{booktabs}       
\usepackage{amsfonts}       
\usepackage{nicefrac}       
\usepackage{microtype}      
\usepackage{lipsum}		
\usepackage{graphicx}
\usepackage{natbib}
\usepackage{doi}

\usepackage{epstopdf}
\usepackage{subfigure}

\usepackage{authblk}

\usepackage{siunitx}
\usepackage{physics}
\AtBeginDocument{\RenewCommandCopy\qty\SI}
\usepackage{algpseudocode}
\usepackage{algorithm}
\usepackage{algorithmicx}
\algnewcommand\algorithmicforeach{\textbf{for each}}
\algdef{S}[FOR]{ForEach}[1]{\algorithmicforeach\ #1\ \algorithmicdo}
\usepackage{siunitx}
\usepackage{threeparttable}
\usepackage{multirow}
\usepackage{siunitx}
\usepackage{arydshln}
\usepackage{pdflscape}
\usepackage[page]{appendix}

\usepackage{comment}
\usepackage{chngpage}

\begin{document}


\title{No-Arbitrage Deep Calibration for Volatility Smile and Skewness}
\author[$\dag$]{Kentaro Hoshisashi$^{\ast}$\thanks{k.hoshisashi@ucl.ac.uk}}
\author[$\dag$]{Carolyn E. Phelan}
\author[$\dag$]{Paolo Barucca}
\affil[$\dag$]{Department of Computer Science \\ University College London \\ Gower Street, London, WC1E 6BT, UK}

\maketitle

\begin{abstract}
Volatility smile and skewness are two key properties of option prices that are represented by the implied volatility (IV) surface. However, IV surface calibration through nonlinear interpolation is a complex problem due to several factors, including limited input data, low liquidity, and noise. Additionally, the calibrated surface must obey the fundamental financial principle of the absence of arbitrage, which can be modeled by various differential inequalities over the partial derivatives of the option price with respect to the expiration time and the strike price. To address these challenges, we have introduced a Derivative-Constrained Neural Network (DCNN), which is an enhancement of a multilayer perceptron (MLP) that incorporates derivatives in the objective function. DCNN allows us to generate a smooth surface and incorporate the no-arbitrage condition thanks to the derivative terms in the loss function. In numerical experiments, we train the model using prices generated with the SABR model to produce smile and skewness parameters. We carry out different settings to examine the stability of the calibrated model under different conditions. The results show that DCNNs improve the interpolation of the implied volatility surface with smile and skewness by integrating the computation of the derivatives, which are necessary and sufficient no-arbitrage conditions. The developed algorithm also offers practitioners an effective tool for understanding expected market dynamics and managing risk associated with volatility smile and skewness.
\end{abstract}


\section{Introduction}
Implied volatility (IV) is the volatility that is hypothesized to make sense of empirical option prices, i.e. implied volatility is the value of volatility which would result in the market price using the Black-Scholes formula. The standard calibration of the premium surface requires us to find a solution to the Black-Scholes partial differential equation (PDE). Traditional calibration mainly involves minimizing the difference between values predicted by a model and those observed, nevertheless no-arbitrage conditions should ensure that the price of any derivative is fixed at the same level as the value of a replicating portfolio, as shown in \cite{delbaen1994general}. Under these conditions, it is a necessary and sufficient condition that certain derivatives of the option prices should satisfy certain derivative inequalities by \cite{carr2005note}. Taking into account the no-arbitrage conditions enhances the robustness and interpretability of the calibration of the premium surface, in particular in presence of sparse market data.


In such a situation, researchers have been investigating how effective the artificial neural networks (ANNs) approach is applied to the calibration problem in the options market, based on its use as a universal approximator as proved in \cite{cybenko1989approximation, hornik1989multilayer, hornik1990universal}. In this study, we expand the standard ANNs backpropagation with derivative terms in the loss function and thus incorporate it into the calibration process, enhancing accuracy. Specifically, this study proposes an expansion of Derivative-Constrained Neural Network (DCNN) described in \cite{lo2023training}, which is the extension of a multilayer perceptron (MLP) with Automatic Differentiation (AD) to compute the exact derivatives simultaneously. This approach ensures the network's differentiability, which is essential for representing derivative functions of the original function of MLP, with an expansion of an MLP with reverse AD that generates the first and second derivatives efficiently in \cite{speelpenning1980compiling}. The resulting network has a deep learning architecture that allows the efficient computation of the derivatives and can therefore introduce differential soft constraints for the generated surface.

To evaluate the representation capability of smile and skewness features in IV surfaces as seen in \cite{rubinstein1985nonparametric, corrado1997implied}, we use one of the stochastic volatility models, the Stochastic Alpha Beta Rho (SABR) model introduced by \cite{hagan2002managing}, often used by practitioners. We utilize sparse option premiums generated using the SABR model and evaluate the ability to reproduce premiums and risk profiles which retain the characteristics of volatility smile and skewness. Through our DCNN network, the interpolation of the premium surface is improved, benefiting from the efficient computation of derivatives and the consideration of no-arbitrage conditions.

An effective model necessitates precise calibration of observable data of option premiums. When there is a model that has explanatory power, it is possible to obtain a more accurate probability distribution using historical data, and it will also be possible to perform derivative evaluation and risk management on the same model. Empirically, it has been observed that for models with small pricing errors, the skewness and kurtosis of the unconditional probability distribution implied by the model under the risk-neutral measure are significantly different from those under historical data, as shown in \cite{feldhutter2016can}. We also demonstrate that this model serves as a powerful tool for practitioners to understand the expected market dynamics and manage risks with volatility smile and skewness.

\section{Background and Literature Review}
The effectiveness of ANNs in addressing function approximation problems has been thoroughly researched across several fields. The fields that bear relevance to this study are detailed below.

\subsection{Volatility smile and skewness} In the early 1970s, the \cite{black1973pricing} model facilitated the pricing of options based on the assumption that the underlying asset's volatility is constant. However, empirical observations revealed that options with different strike prices actually implied varied volatilities, known as the volatility smile and skewness, as shown in \cite{corrado1997implied} and \cite{rubinstein1985nonparametric}. Numerous local volatility models, such as \cite{dupire1994pricing} and \cite{derman1996local}, attempted to account for the reproduction of the static pattern of the smiles. Effective prediction of their dynamics was made possible by the introduction of stochastic volatility models, such as \cite{hull1990pricing, heston1993closed, hagan2002managing}, and \cite{gatheral2004parsimonious}. One common practical approach is the SABR stochastic volatility model by \cite{hagan2002managing}, which employs parameters to characterize the smile and skewness in its stochastic differential equations. However, the dynamics of the volatility smile are still not perfectly represented, necessitating frequent recalibration to align the model with market data. Related to these challenges, the potential of ANNs for solving PDEs has been explored in \cite{barhak2001parameterization,chen2018neural,khoo2021solving}.

\subsection{Financial applications of ANNs} Previous research into finding option premiums using ANNs has primarily focused on IV and has often employed adapted global optimization methods, such as \cite{liu2019neural, cont2022simulation, choudhary2023funvol}. Additionally, there have been applications of advanced network models, such as variational autoencoders in \cite{bergeron2022variational}, generative Bayesian models in \cite{jang2019generative}, hybrid gated NN 
in \cite{cao2021option}, and Vol GANs in \cite{cont2022simulation}. Several studies have addressed the calibration problems of option products by penalizing the loss using soft constraints, as shown in \cite{itkin2019deep, ackerer2020deep,choudhary2023funvol, cont2022simulation} or mapping pricing from model parameters, such as \cite{bayer2019deep, mcghee2020artificial, horvath2021deep}. A comprehensive review of ANNs methods for option pricing was conducted in \cite{ruf2019neural}.

\subsection{Multi-task deep learning} Multi-task deep learning described in \cite{zhang2014improving, strezoski2017omniart} and multi-objective optimization 
in \cite{gunantara2018review} address the challenge of weighing multiple loss functions to obtain better performance in \cite{kendall2018multi, marquez2017imposing} with both soft and/or hard constraints for ANNs.
In this context, the introduction of derivative terms in the loss function to fit full PDEs has been explored in physics-informed neural networks (PINNs) introduced by \cite{raissi2019physics}, and DCNNs in \cite{lo2023training} where isolated derivative terms are considered by \cite{yeh2010first, yao2020pyhessian, pizarroso2020neuralsens}.

\section{Calibration Problem}
The calibration of volatility surfaces or option prices is an important inverse problem in quantitative finance. In \cite{dupire1994pricing}, the author has proposed the local volatility model, in which the European options prices satisfy the PDE of the following form with the current price of an underlying asset $S_t>0$,
\begin{equation}
-\frac{1}{2} \sigma^2(K, \tau) K^2 \frac{\partial^2 C}{\partial K^2}+\frac{\partial C}{\partial \tau}+ (r-q) K \frac{\partial C}{\partial K}+qC=0,
\end{equation}
with initial and boundary conditions given by
\begin{equation}
\begin{aligned}
    C(K, 0) &=\left(S_T-K\right)^{+}, \\
    \lim _{K \rightarrow \infty} C&(K, \tau) =0, \\
    \lim _{K \rightarrow 0} C&(K, \tau) =S_t,
\end{aligned}
\label{eq: initial and boundary conditions}
\end{equation}
where $K$ is the strike price and $\tau$ is the option term to expiry $T$ from valuation date $t$, and $C=C(K, \tau)$ is the value of the European call option with expiration date $T$, strike price $K$ with $K, \tau \in [0, \infty)$, a risk-free rate $r$, and a dividend yield $q$. The inverse problem of the implied volatility model is that, given limited options prices, we would like to know the premium function $C(K, \tau)$ or an implied (not local) volatility surface, which gives these options prices via the Black-Scholes formula. The challenge of this inverse problem is the scarcity of options price data. To solve this, possibilities are to interpolate/extrapolate the price data or add further information relevant to the problem.

\subsection{No-arbitrage constraints of European options}
\label{section: No-arbitrage constraints of European Options}
The calibration of option prices is limited by sparse data and, as discussed, should obey the constraints imposed by the no-arbitrage conditions. The no-arbitrage principle posits that market prices prevent guaranteed returns above the risk-free rate. We consider the necessary and sufficient conditions for no-arbitrage, reported in \cite{carr2005note}. In fact, these conditions had been implemented by \cite{ait2003nonparametric} and were later also used by \cite{roper2010arbitrage, fengler2015semi} as the constraints that our surface needs to satisfy. This allows us to appropriately express the call option price as a two-dimensional surface. The necessary and sufficient conditions for no-arbitrage are represented as non-strict inequalities for several first and second derivatives,
\begin{equation}
    -e^{-r\tau} \leq \pdv{C}{K}\leq 0, \;\;\; \pdv[2]{C}{K} \geq 0, \;\;\; \pdv{C}{\tau}\geq 0.
    \label{eq: no-arbitrage conditions}
\end{equation}
In Eq.~\eqref{eq: no-arbitrage conditions}, no-arbitrage conditions require these derivatives to have a specific sign. The standard architecture does not automatically satisfy these conditions when calibrating with a loss function simply based on the mean squared error (MSE) for the prices.

\subsection{ANNs with No-arbitrage Constraints}
Here, we can consider the problem as a calibration problem of an approximated price function defined by the ANNs, $\hat{\Phi}(K, \tau)$. Having obtained each derivative constraint from $\hat{\Phi}$, we can now introduce the total cost function with no-arbitrage constraints for European (call) options to minimise,
\begin{equation}
    E(\mathbf{C}, \hat{\Phi}) = E_{{\text{{\scriptsize MSE}}}} + E_\mathcal{P},
\end{equation}
\begin{equation}
    E_{\text{{\scriptsize MSE}}} = \frac{1}{N} \sum_{i=1}^N \left\{C(K_i, \tau_i) -\hat{\Phi}(K_i, \tau_i) \right\}^2.
    \label{eq: E_MSE}
\end{equation}
Here, $C(K_i, \tau_i)$ is the observed premium for the indexed values of strike $K_i$ and time to expiry $\tau_i$, $i=1,\ldots, N$ from the observed dataset. The penalty term $E_\mathcal{P}$ represents a score of the total no-arbitrage conditions involving the first and second derivatives,
\begin{multline}
    E_\mathcal{P} = \frac{1}{M^{(K)}M^{(\tau)}} \sum_{i=1}^{M^{(K)}} \sum_{j=1}^{M^{(\tau)}} \left\{ \lambda \left(m_{\text{K}}, \pdv{\hat{\Phi}(\hat{K}_j,\hat{\tau}_j)}{\hat{K}_j} \right) + \lambda \left(m_{\text{KK}}, \pdv[2]{\hat{\Phi}(\hat{K}_j,\hat{\tau}_j)}{\hat{K}_j} \right) \right.\\
    \left. + \lambda \left(m_{\tau}, \pdv{\hat{\Phi}(\hat{K}_j,\hat{\tau}_j)}{\hat{\tau}_j} \right) \right\}.
    \label{eq: E_p}
\end{multline}
The set of $\hat{K}$ and $\hat{\tau}$ values in which we evaluate the derivatives can be obtained with a mesh grid, $j=1,\ldots, M$. The $h$ terms are sign-adjustment coefficients, which make sure that the signs of the penalty terms are correct, $h_K = 1, h_{KK} = -1, h_\tau = -1$. $\lambda(m,x)$ is an intensifier of derivative losses in the total cost,
\begin{equation}
    \lambda(m,x) = \begin{cases}m\cdot g(\mathbf{x}), & \text{if penalty} \\ 0, & \text{if not penalty}\end{cases}
    \label{eq: lambda function in loss}
\end{equation}
where $m \in \mathbb{R}$ are constants and $g(x)$ are intensifier functions, which can be non-linear.

\section{Derivative-Constrained Neural Network (DCNN)}
\label{section: Derivative-Constrained Neural Network}
This section introduces the Derivative-Constrained Neural Network (DCNN) as an expansion of the work in \cite{lo2023training}, which efficiently computes the partial derivatives of a neural network function with respect to its input features. We consider a simple feed-forward neural network architecture, a multilayer perceptron (MLP). Let $L \geq 2$ be an integer representing the depth of the network; we consider a neural network constructed with one input vector, $L$ hidden layers, and one output value. Both the input values and the output variable are real numbers, i.e., $\mathbf{x} \in \mathbb{R}^n$ and $y \in \mathbb{R}$. We can consider the MLP as a multivariate function $\Phi$ depending on the variables  $\mathbf{x}$, i.e. $\Phi:\mathbf{x} \mapsto y$,
\begin{align}
    y = \Phi(\mathbf{x}) = A_L \circ f_{L-1} \circ A_{L-1} \circ \cdots \circ f_1 \circ A_1(\mathbf{x}),
    \label{eq: MLP generative process}
\end{align}
where for $l=1,\ldots , L$, $A_{l}: \mathbb{R}^{d_{l-1}} \rightarrow \mathbb{R}^{d_{l}}$ are affine functions as $A_l(\mathbf{x}_{l-1}) = W_l^\mathsf{T} \mathbf{x}_{l-1} + \mathbf{b}_l$, and $d_l$ is the number of neurons in the next layer $l$ for $\mathbf{x}_{l-1} \in \mathbb{R}^{d_{l-1}}$, with $W_l \in \mathbb{R}^{d_{l-1} \times d_l}$ and $\mathbf{b}_l \in \mathbb{R}^{d_l}$,  $d_{0}=n$, $d_{L}=1$, and $\mathbf{x}_0=\mathbf{x}$. $f_l$ is an activation function which is applied component-wise. Given a dataset $\mathbf{X}$, which includes a set of pairs $(\mathbf{x}^{(i)}, y^{(i)}), i = 1,\ldots N$, and a cost function $E(\mathbf{X}, \Phi)$, the network model $\Phi$ is found by fitting the values of $W$ and  $\mathbf{b}$ which minimize the cost function.

We consider optimization problems which include losses, not only associated with the network function $\Phi(\mathbf{x})$, but also its derivatives. The total cost includes several terms, which account for the derivatives, 
\begin{equation}
\begin{split}
    E(\mathbf{X}, & \hat{\mathbf{X}}, \Phi) := \frac{1}{N} \sum_{i=1}^N \left( y^{(i)} - \Phi(\mathbf{x}^{(i)}) \right)^2 \\
    &+ \frac{1}{M} \sum_{j=1}^M \left\{ \lambda_{m_1} ( h_1 \nabla\Phi(\hat{\mathbf{x}}^{(j)})) + \lambda_{m_2} ( h_2 {\nabla}^{2}\Phi(\hat{\mathbf{x}}^{(j)})) \right\},
    \label{eq: cost function}
\end{split}
\end{equation}
where, $\hat{\mathbf{x}}^{(j)}$ is input features for derivative losses in computational grids $\hat{\mathbf{X}}$ for $j=1,\ldots , M$, and $h$ are vectored signs. $\nabla$ and $\nabla^{2}$ are partial derivative vectors respectively,
\begin{equation}
\small
    \nabla\Phi(\mathbf{x}) = \left[ \pdv{\Phi}{x_1}, \pdv{\Phi}{x_2}, \ldots, \pdv{\Phi}{x_n} \right]^\top, 
    \nabla^2\Phi(\mathbf{x}) = \left[ \pdv[2]{\Phi}{x_1}, \pdv[2]{\Phi}{x_2}, \ldots, \pdv[2]{\Phi}{x_n} \right]^\top.
    \label{eq: definition of nabla and nabla squeared}
\end{equation}

\begin{figure}[htbp]
    \centering
    \includegraphics[width=0.85\columnwidth]{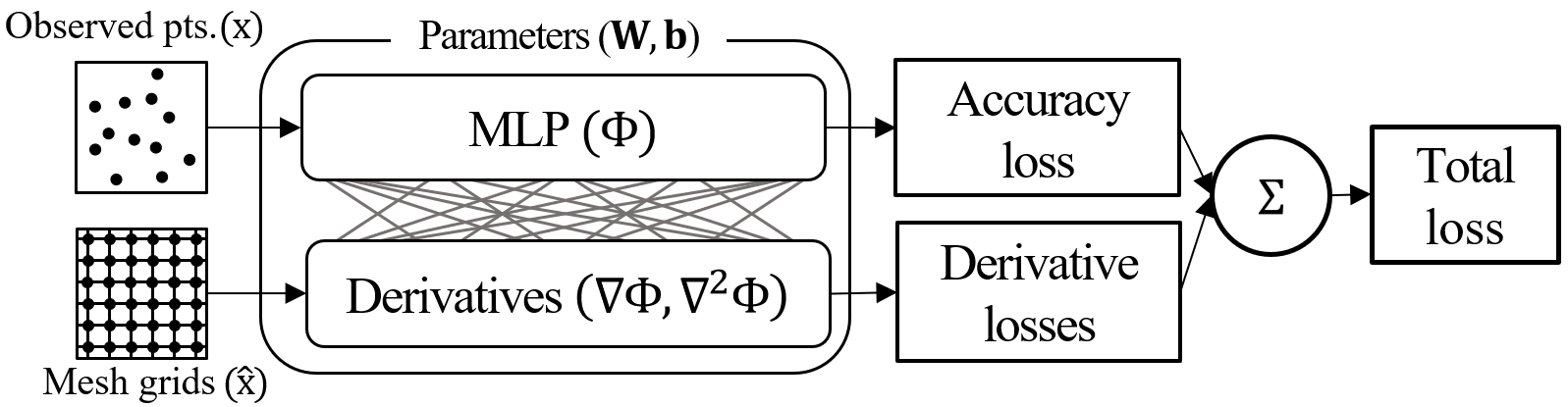}
    \caption{The whole network architecture of Derivative-Constrained Neural Network (DCNN). It is an expansion of a multilayer perceptron (MLP) with additional networks that simultaneously generate its first and second derivatives.}
    \label{Figure: NoArbSS_DCNN_Architecture}
\end{figure}
Using Eq.~\eqref{eq: cost function} as the cost function, the optimizer uses gradients with respect to the parameters (i.e., $W, \mathbf{b}$) for updates. A challenge lies in that Eq.~\eqref{eq: cost function} involves derivatives with respect to $\mathbf{x}$, also functions of the parameters. When numerical approximation of derivatives is used, it could result in slow or inaccurate solutions. To solve this, this study utilizes DCNN, an extended backpropagation algorithm in \cite{rumelhart1986learning} with Automatic Differentiation, for the gradient in Eq.~\eqref{eq: cost function} through exact derivative formulations. Following \cite{yeh2010first} and \cite{pizarroso2020neuralsens}, the first derivatives of a layer in $\Phi$ is
\begin{equation}
\nabla_l := \pdv{\mathbf{x}_{l}}{\mathbf{x}} 
= \left\{ f_l^\prime \circ A_l \left(\mathbf{x}_{l-1}\right) \ast W_{l}^\top \right\}\nabla_{l-1},
\label{eq: nabla layer}
\end{equation}
with $\ast$ denoting tensor broadcasting in \cite{van2011numpy}. We then compute the first derivative of $\Phi$, i.e., $\nabla \Phi$, by sequentially applying the chain rule in Eq.~\eqref{eq: nabla layer}. Note, $f^\prime$ of the last layer and $\nabla_{l-1}$ of the first layer aren't applied in Eq.~\eqref{eq: nabla layer}.
For the second derivative of $\Phi$, we use the definition $\partial^2 \Phi / \partial \ x^2 = \partial (\partial \Phi / \partial x) / \partial x$. The second derivative of each layer is described as 
\begin{equation}
\begin{split}
    \nabla^2_l 
    := \pdv[2]{\mathbf{x}_{l}}{\mathbf{x}} 
    = f_l^{\prime\prime}\circ & A_l\left(\mathbf{x}_{l-1}\right) \ast \left\{W_{l}^\top \nabla_{l-1}\right\}^{\circ 2} \\ 
    &+\left\{ f_l^\prime \circ A_l \left(\mathbf{x}_{l-1}\right) \ast W_{l}^\top \right\} \nabla^2_{l-1}.
    \label{eq: nabla squared layer}
\end{split}
\end{equation}
and $\left\{\cdot\right\}^{\circ 2}$ is the operation of the Hadamard product for element-wise as described in \cite{reams1999hadamard}. Subsequently, $\nabla^2\Phi(\mathbf{x}$) is obtained using Eqs.~\eqref{eq: nabla layer} and \eqref{eq: nabla squared layer} with the chain rule. These formulations require MLP activation functions to be second-order differentiable or higher. It is noted that functions like Relu or Elu need slight additional consideration at non-differentiable singular points. If all activation functions are second-order differentiable, the same is true for the whole network as shown in \cite{hornik1990universal}.

\subsection{Algorithms}

The Derivative-Constrained Neural Network (DCNN) algorithm efficiently computes the partial derivatives of a neural network function with respect to its input features.
\begin{algorithm}
\centering
\caption{Derivative-Constrained Neural Network Algorithm}\label{alg: DCNN}
\begin{algorithmic}[1]
    \Require Training dataset of size $N$ $\mathbf{X}:~(\mathbf{x}^{(i)}, y^{(i)}), i = 1,\ldots N$ \\
    \hspace{24pt}Mesh grid points of size $M$ $\hat{\mathbf{X}}:~\hat{\mathbf{x}}^{(j)}, j = 1,\ldots M$ \\
    \hspace{24pt}An MLP $\Phi$ with $L$ layers, $l = 1,\ldots L$, weights $W$, bias $b$ \\
    \hspace{24pt}Cost function $E(\mathbf{X}, \hat{\mathbf{X}}, \Phi_{W,\mathbf{b}})$ \\
    \hspace{24pt}Derivative function by backpropagation $\delta$ \\
    \hspace{24pt}Optimizer function of the gradient $g(x)$ \\
    \hspace{24pt}Max number of epochs $I_{\text{max}}$
    \State $\text{Initialize}~~ W, \mathbf{b} \leftarrow \text{random numbers}$
    \For{$k \leftarrow 1 ~ \text{to} ~ {I_{\text{max}}}$}
        \State \text{/* Propagate forward MLP in training dataset */}
        \ForAll{\text{training data} $\left(\mathbf{x}, y\right)\in\mathbf{X}$}
            \State $\mathbf{e}_1 \leftarrow y-\Phi_{W,\mathbf{b}}\left(\mathbf{x}\right)$
        \EndFor
        \State \text{/* Calculate the derivatives of MLP in mesh grid points */}
        \ForAll{\text{mesh grid point} $\hat{\mathbf{x}}\in\hat{\mathbf{X}}$}
            \State $\mathbf{e}_2, \mathbf{e}_3 \leftarrow \nabla\Phi(\hat{\mathbf{x}}), \nabla^2\Phi(\hat{\mathbf{x}})$
        \EndFor
        \State \text{/* Calculate the total cost */}
        \State $\mathbf{e}_4 \leftarrow E(\mathbf{X}, \hat{\mathbf{X}}, \Phi_{W,\mathbf{b}})$
        \State \text{/* Calculate partial derivatives with respect to parameters}
        \State \text{\hspace{55pt} of the cost function by backpropagation */}
        \State $\pdv{E}{W}, \pdv{E}{\mathbf{b}} \leftarrow \delta_{W}E, \delta_{\mathbf{b}}E$
        \State \text{/* Update the weights and bias */}
        \State $W,\mathbf{b} \leftarrow W-g\left(\pdv{E}{W}\right), \mathbf{b}-g\left(\pdv{E}{\mathbf{b}}\right)$
    \EndFor
    \State \textbf{return} $\Phi$, $\mathbf{e}$
\end{algorithmic}
\end{algorithm}
Algorithm \ref{alg: DCNN} exhibits characteristics that set it apart from conventional learning methods. First, the computation points $\hat{\mathbf{X}}$ for the derivatives of the MLP do not correspond with the points of the training dataset $\mathbf{X}$, which is typically sparse and unbalanced. The algorithm adjusts the derivatives to fit mesh grids, hence capturing derivative data across a wide array of input features. Secondly, the cost function $E$ does not depend only on the MLP's direct output but also on its derivatives as specified in Eq.\eqref{eq: cost function}, all of which depend on identical network parameters. DCNN facilitates accurate calibration and gradient computation of the parameters through precise formulations, which consist of a linear transformation and the activation function's derivatives, as described in the previous section.

\section{Testing with Synthetic Data}
The developed algorithm (i.e. DCNN) for evaluating volatility smile and skewness was first tested on simulated values in a parameterized two-dimensional case of the surface interpolation problem. We took up the well-known Stochastic Alpha Beta Rho (SABR) model introduced by \cite{hagan2002managing} and prepared a sparse two-dimensional dataset to test our methodology. As an empirical experiment, we applied DCNN for the surface interpolation with real market data, which are sparse, and examined the efficiency of the solution for the surface with no-arbitrage constraints.

\subsection{The SABR model}
The SABR model in \cite{hagan2002managing} is a typical parametric model, which can capture the market volatility smile and skewness and reasonably depict market structure. When $F_t$ is defined as the forward price of an underlying asset at time $t$, the SABR model is described as
\begin{equation}
\begin{split}
dF_t&=\alpha_t F_t^\beta dW_t^1,\\
d\alpha_t&=\nu\alpha_t\ dW_t^2,\\
\langle dW_t^1 & , dW_t^2 \rangle=\rho dt.
\end{split}
\end{equation}
Here, $W_t^1$, $W_t^2$ are standard Wiener processes, $\alpha_t$ is the model volatility, $\rho$ is the correlation between the two processes, and $\nu$ is analogous to vol of vol in the Heston model defined in \cite{heston1993closed}, these are parameters corresponding to skewness and smiles. The additional parameter $\beta$  describes the slope of the skewness. A significant feature of the SABR model is that the price of the European option can be formulated in closed form, as shown in \cite{hagan2002managing}, up to the accuracy of a series expansion. Essentially, it is shown there that the IV in the SABR model is given by the appropriate formula in Eq (16) from \cite{black1976pricing}. For given $\alpha, \beta, \rho, v$ and $F=S_t e^{r \tau}$ with a fixed risk-free rate under the risk-neutral measure in \cite{hull1993options}, this volatility is given by:
\begin{equation}
\begin{split}
    \sigma_{\text{{\scriptsize SABR}}}(K, \tau)
    &=\frac{\alpha\left(1+\left(\frac{(1-\beta)^2}{24} \frac{\alpha^2}{(F K)^{1-\beta}}+\frac{1}{4} \frac{\rho \beta v \alpha}{(F K)^{(1-\beta) / 2}}+\frac{2-3 \rho^2}{24} v^2\right) \tau\right)}{(F K)^{(1-\beta) / 2}\left[1+\frac{(1-\beta)^2}{24} \ln ^2 \frac{F}{K}+\frac{(1-\beta)^4}{1920} \ln ^4 \frac{F}{K}\right]} \frac{z}{\chi(z)}, \\
    z=&\frac{v}{\alpha}(F K)^{(1-\beta) / 2} \ln \frac{F}{K}, \; \chi(z)=\ln \left(\frac{\sqrt{1-2 \rho z+z^2}+z-\rho}{1-\rho}\right).
\end{split}
\label{eq: SABR formula}
\end{equation}
Note as in \cite{hagan2002managing} that if $K=F$ then the $z$ and $\chi(z)$ terms are removed from the equation, as then $\frac{z}{\chi(z)}=1$ in the sense of a limit, and so
\begin{equation}
\sigma_{\text{{\scriptsize SABR}}}(F, \tau)=\frac{\alpha\left(1+\left(\frac{(1-\beta)^2}{24} \frac{\alpha^2}{F^{2-2 \beta}}+\frac{1}{4} \frac{\rho \beta v \alpha}{F^{1-\beta}}+\frac{2-3 \rho^2}{24} v^2\right) \tau\right)}{F^{1-\beta}}.
\label{eq: SABR formula ATM}
\end{equation}
Discussion and analysis of parameters setting methodologies for the SABR model with limited input data are also discussed in \cite{west2005calibration}. 

Finally, option premiums for the experiment are computed from the volatility $\sigma_{\text{{\scriptsize SABR}}}$ by the Black formula introduced by \cite{black1976pricing}. This is similar to the Black-Scholes formula by \cite{black1973pricing} for valuing options, except that the underlying spot price is replaced by a discounted forward price $F$. Then, the Black formula states the price for a European call option is
\begin{equation}
C(K, \tau)=e^{-r \tau}\left[F N\left(d_1\right)-K N\left(d_2\right)\right],
\end{equation}
where
\begin{equation}
d_1 = \frac{\ln (F / K)+\left(\sigma_{\text{{\scriptsize SABR}}}^2 / 2\right) \tau}{\sigma \sqrt{\tau}}, \; d_2= d_1-\sigma_{\text{{\scriptsize SABR}}} \sqrt{\tau},
\end{equation}
and $N(\cdot)$ is the cumulative normal distribution function.

\begin{figure}[htbp]
    \centering
    \includegraphics[width=0.6\columnwidth]{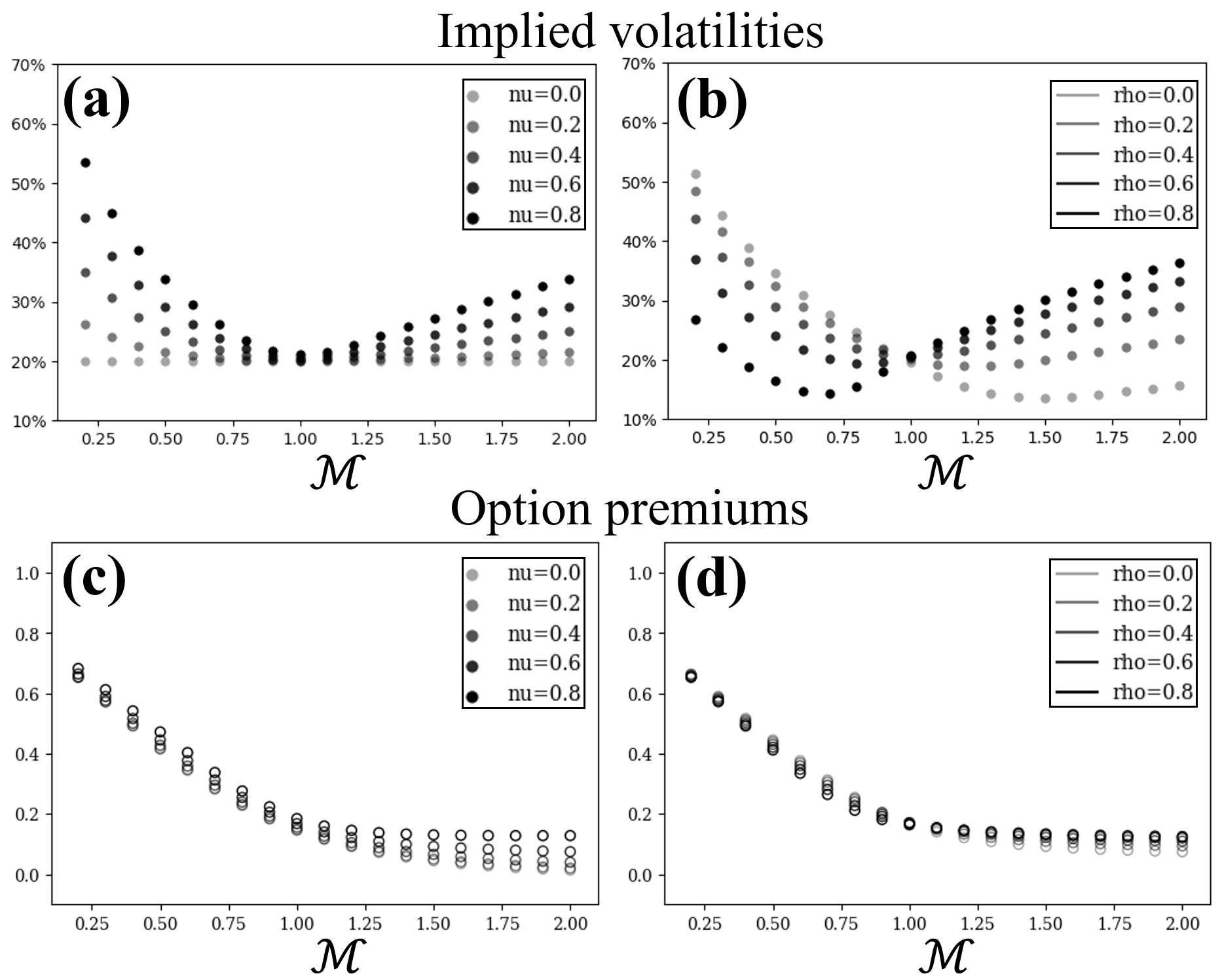}
    \caption{Sparse data of implied volatility (IV) and option (call) premiums using the Stochastic Alpha Beta Rho (SABR) model by moneyness $\mathcal{M}=K/F$. \textbf{(a)} The volatilities by smile parameters ($\nu$) with fixed $\rho=0$. \textbf{(b)} By skewness parameters ($\rho$) with fixed $\nu=0.6$. \textbf{(c-d)} shows the option (call) premiums with corresponding to \textbf{(a-b)}.}
    \label{Figure: NoArbSS_SABR_smile_skewness_samples}
\end{figure}
It is noted that option characteristics are often encapsulated in IV, but the market variables are option premiums. Figure \ref{Figure: NoArbSS_SABR_smile_skewness_samples} shows the volatilities and premiums under fixed parameters by $F=1, \alpha=0.2$, $\beta=1$, $r=0.04$, $\tau=1$, and $\mathcal{M}=K/F$. By changing smile and skewness parameters, we observe a smaller difference in the numerical values on a premium basis in Figure \ref{Figure: NoArbSS_SABR_smile_skewness_samples} \textbf{(c-d)} than that of volatility in Figure \ref{Figure: NoArbSS_SABR_smile_skewness_samples} \textbf{(a-b)}. Therefore, the calibration task in the actual situation becomes a challenging problem to capture smile and skewness characteristics based on the observed premiums.

\subsection{Experimental design}
\label{subsection: Experimental design}
To investigate the ability of ANNs to recreate the volatility smile and skewness, we prepared sparse option premiums for in-sample testing and dense ones for out-of-sample testing based on the SABR model. Sparse option premiums used for in-sample testing are calculated using mesh grids by 25 points for moneyness $\mathcal{M}=K/F\in[0.1, 2.5]$ and 7 for $T\in[0.1, 0.5, 1, 2, 3, 4, 5]$ with varying smile ($\nu$) and skewness ($\rho$) parameters. Additionally, samples contain the boundary conditions with 200 additional points with $\mathcal{M}\in[0, 2.5]$ and $\tau=0$ or in $T\in[0, 5]$ and $K=0$. After that, these premiums are fitted into various models, such as the cubic spline method, MLP, Arbitrage-free smoothing (AFS) by \cite{fengler2009arbitrage}, and DCNN. For comparison, we evaluate the errors from the predictions using dense mesh girds as out-of-sample, which are equally distributed by 126 points in $\mathcal{M}\in[0, 2.5]$ and 101 points in $T\in[0, 5]$. We use an evaluation metric as followability of smile and skewness as similar to Eq \eqref{eq: E_MSE},
\begin{equation}
    E_{\text{{\scriptsize MSE}}}^{(\sigma)} = \frac{1}{N} \sum_{i=1}^N \left\{ \sigma_{\text{{\scriptsize SABR}}}\left(K_i, \tau_i\right) - \sigma_{\text{{\scriptsize Black}}}\left(\hat{\Phi}\left(K_i, \tau_i\right),F , K_i, r, \tau_i\right) \right\}^2,
    \label{eq: sigma_MSE}
\end{equation}
where $\sigma_{\text{{\scriptsize Black}}}(\cdot)$ is the function which computes IV based on the Black model corresponding to the predicted premium $\hat{\Phi}(K_i, \tau_i)$. To fit a more realistic market situation, the following experiment utilized sparse (and uneven) grid data referring to actual historical traded grids from 10th to 14th July 2023 and generates synthetic option premiums on the grids by Eq. \ref{eq: SABR formula} as in-sample training data.

The base MLP architecture used in this study consists of two fully connected hidden layers with Softplus activation functions, whose derivatives are {\small $f^{\prime}:{\left(1+e^{-x}\right)}^{-1}$, $f^{\prime\prime}:{e^{-x}}{(1+e^{-x})^{-2}}$} in Eqs.~\eqref{eq: nabla layer} and \eqref{eq: nabla squared layer}, with an output layer with a linear function (i.e. no activation function). Each layer has 16 neurons. We set the number of epochs as $10,000$, and Adam, introduced by \cite{kingma2014adam}, as the optimizer using gradient-based training with normally randomised weight initialization. In the cubic spline model, we use the \texttt{SmoothBivariateSpline} function with three degrees in the Scipy package (\cite{virtanen2020scipy}) in Python. Our numerical experiments were run using Pytorch (\cite{NEURIPS2019_9015}) and JAX (\cite{jax2018github}) packages for efficient automatic differentiation on Google Colaboratory (\cite{googlecolab}) with 36 GB of RAM and a dual-core CPU of 2.3 GHz. We consistently used the same random seeds across different conditions for the statistical analysis and changed these seed values ten times. We set $m_1=m_3=0.001, m_2=0.01, g(x)=x$ in Eq. \eqref{eq: E_p}, and $\alpha=0.2$, $\beta=1.0$, $q=0$, and $r=0.04$ as the parameters of the SABR model. In the implementation of AFS proposed in \cite{fengler2009arbitrage}, we utilized the Matlab codes provided by the authors for our calculations. Note that AFS is available for a rectangular mesh grid, although empirical data has scattered grids, and does not define interpolation along the $\tau$ dimension in the paper. However, we adapted \texttt{presmoother} function from their codebase for interpolation along $\tau$, enabling implementation on our rectangular mesh grid per AFS requirements by eliminating grids at $\tau=0$ and reducing grids at $K=0$ compared to other models. The visualizations in the figures of the results section were based on the model outputs when $\nu=0.6$ and $\rho=-0.4$.

\subsection{The S\&P 500 dataset}
The empirical experiments were conducted using intraday prices for S\&P 500 call options from 10th to 14th July 2023; we obtained about two thousand points on a daily basis via the Yahoo finance library as Table \ref{table: SP500 options}. We added the synthesized points corresponding to boundary conditions to training, and all other setups are the same as in the previous section.
We also conducted the backtests summarized in Appendix \ref{appendix: Backtests}.

\begin{table}
\begin{center}
\caption{A summary of intraday prices for S\&P 500 options.}
{\small\begin{tabular}{@{}lccccc}
    \toprule
        & \small{10-Jul-23} & \small{11-Jul-23} & \small{12-Jul-23} & \small{13-Jul-23} & \small{14-Jul-23} \\
    \midrule
        count & 2,066 & 1,930 & 2,261 & 1,757 & 1,963 \\
        $\tau_{\rm{min}}$ & 0.005 & 0.000 & 0.000 & 0.000 & 0.000 \\
        $\tau_{\rm{max}}$ & 5.444 & 4.438 & 5.433 & 5.430 & 5.427 \\
        $\mathcal{M}_{\rm{min}}$ & 0.022 & 0.022 & 0.041 & 0.043 & 0.042 \\
        $\mathcal{M}_{\rm{max}}$ & 2.312 & 2.310 & 2.297 & 1.939 & 1.926 \\
    \bottomrule
\end{tabular}}
\label{table: SP500 options}
\end{center}
\end{table}

\subsection{Results}
We obtained sparse in-sample data for training and dense out-of-sample data to compare the effectiveness of interpolation with volatility smile and skewness by models: cubic spline, MLP, and DCNN.

\begin{figure}[H]
    \centering
    \includegraphics[width=0.6\columnwidth]{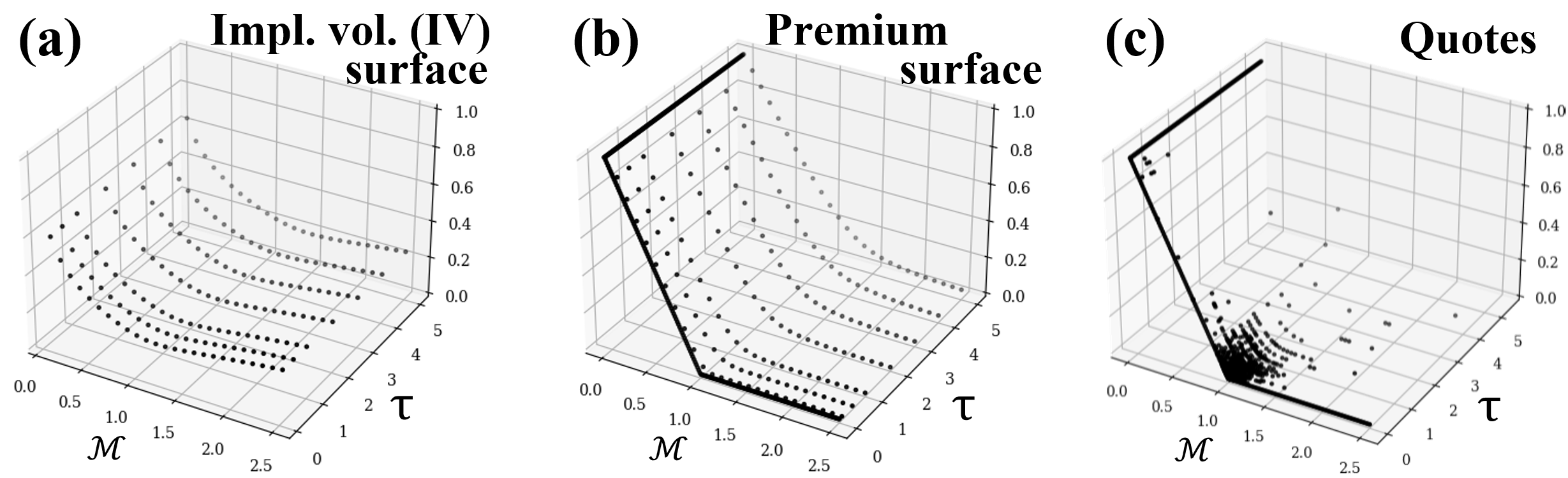}
    \caption{Sparse samples for the training set. \textbf{(a)} The IVs with the fixed SABR parameter ($\nu=0.6, \rho=-0.4$). \textbf{(b)} the option premiums corresponding to \textbf{(a)} with initial and boundary conditions in Eq.\eqref{eq: initial and boundary conditions}. \textbf{(c)} S\&P 500 option premiums traded intra-day as of 12th July 2023, with boundary conditions.} 
    \label{Figure: NoArbSS_inSamples}
\end{figure}
At first, we show the issue associated with generating the IV surface near boundary conditions. As shown in \ref{Figure: NoArbSS_inSamples} \textbf{(a)}, we can prepare the whole volatility surface using the SABR model; however, the original option premiums in Figure \ref{Figure: NoArbSS_inSamples} \textbf{(c)} largely comprise of at-the-money options with a short time to expiry which makes it is hard to identify a complete implied volatility surface from premiums.

\begin{figure}[htbp]
    \centering
    \includegraphics[width=0.85\columnwidth]{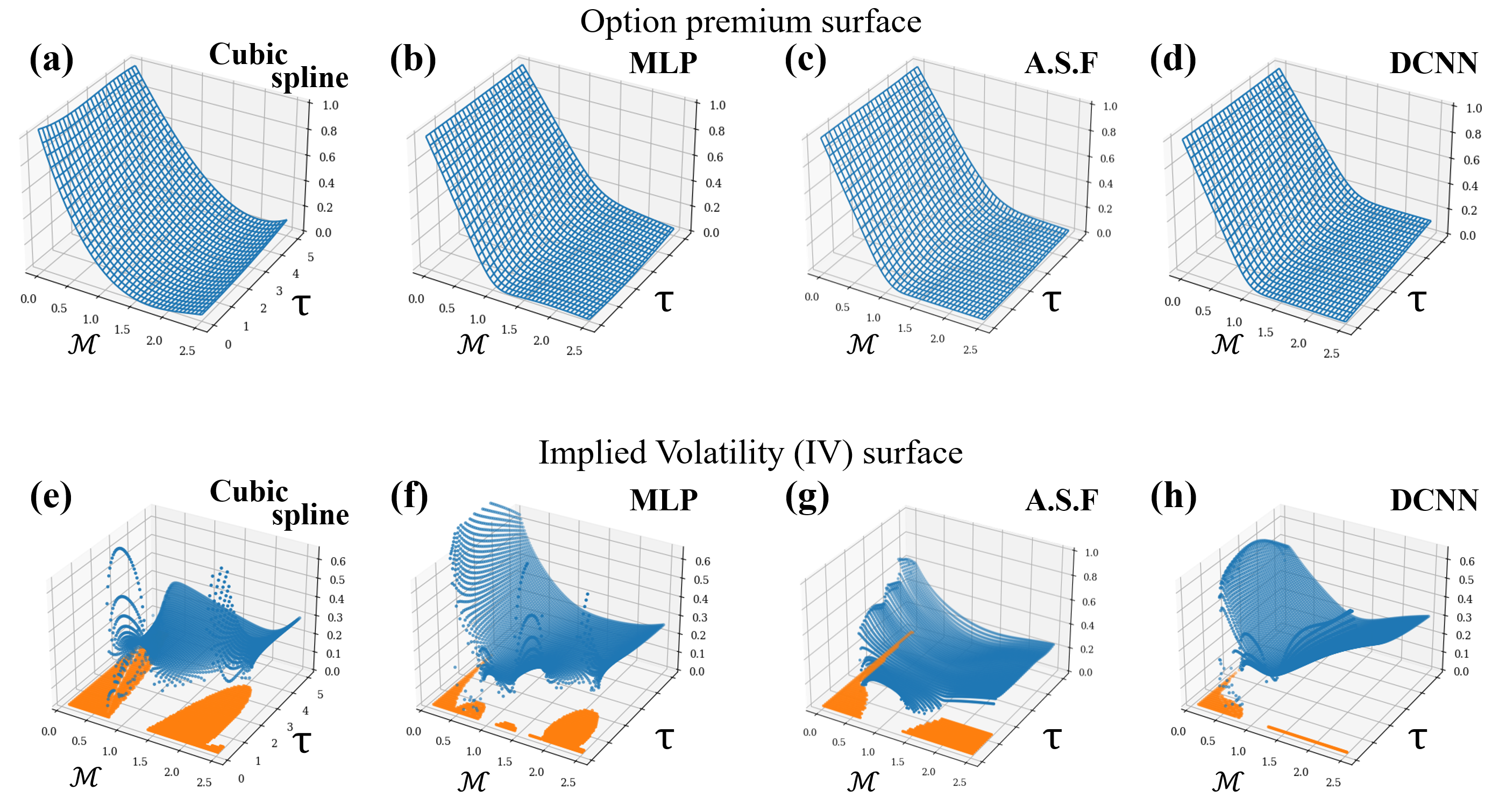}
    \caption{Calibrated models fitted by synthesized option premiums via SABR model \textbf{(a-d)}, and IV surface completed by those premiums \textbf{(e-h)}. Orange points show the invalid IVs computed by corresponding to option premiums \textbf{(a-c)}.}
    \label{Figure: NoArbSS_prem_iv_vals}
\end{figure}
Figure \ref{Figure: NoArbSS_prem_iv_vals} \textbf{(a-d)} illustrates the predicted results for option premiums from trained models derived from out-of-sample data. In Figure \ref{Figure: NoArbSS_prem_iv_vals}\textbf{(e-f)}, we compare the profiles of the IV surface created by the cubic spline method, MLP, AFS, and DCNN. Training times were approximately 30 seconds for a simple MLP and 50 for DCNN. Note again that calibration provided option premium surfaces, which were converted to the implied volatility surface. As a result, the cubic spline method, a simple interpolation approach, reveals the largest area of invalid volatility, signifying a higher IV error. In contrast, more sophisticated techniques, like AFS and particularly DCNN, display fewer errors. Additionally, DCNN gives the most stable results for IV, indicating that it effectively generates a surface that can represent smile and skewness features, thus characterizing the IV surface more precisely.

\begin{table}
\begin{center}
\begin{minipage}{170mm}
    \caption{Premium, penalty and volatility errors in out-of-sample. \textbf{Bold values} indicate lower (better) values among the models.}
    {\small\begin{tabular}{@{}ccrclrclrclrcl}
    \toprule
    \multicolumn{2}{c}{\text {Conditions}} & \multicolumn{12}{c}{\text {Out-of-sample Errors }} \\
    \multicolumn{2}{c}{\text {SABR param.}} & \multicolumn{3}{c}{\text {Cubic Spline}} & \multicolumn{3}{c}{\text {MLP}} &\multicolumn{3}{c}{\text {AFS}$^{\rm a}$} &\multicolumn{3}{c}{\text {DCNN}} \\
    ~~~~~\multirow{2}{*}{$\nu$} & \multirow{2}{*}{$\rho$} &\vspace{3pt}  $E_{\text{{\scriptsize MSE}}}$ & $E_\mathcal{P}$ & $E_{\text{{\scriptsize MSE}}}^{(\sigma)}$ & $E_{\text{{\scriptsize MSE}}}$ & $E_\mathcal{P}$ & $E_{\text{{\scriptsize MSE}}}^{(\sigma)}$ & $E_{\text{{\scriptsize MSE}}}$ & $E_\mathcal{P}$ & $E_{\text{{\scriptsize MSE}}}^{(\sigma)}$ & $E_{\text{{\scriptsize MSE}}}$ & $E_\mathcal{P}$ & $E_{\text{{\scriptsize MSE}}}^{(\sigma)}$ \\
    & & {\scriptsize $\times 10^{-4}$} & {\scriptsize $\times 10^{-3}$} & {\scriptsize $\times 10^{-2}$} & {\scriptsize $\times 10^{-4}$} & {\scriptsize $\times 10^{-3}$} & {\scriptsize $\times 10^{-2}$} & {\scriptsize $\times 10^{-4}$} & {\scriptsize $\times 10^{-3}$} & {\scriptsize $\times 10^{-2}$} & {\scriptsize $\times 10^{-4}$} & {\scriptsize $\times 10^{-3}$} & {\scriptsize $\times 10^{-2}$} \\
    \midrule
    ~~~~~0.0&0.0&6.25&10.65&1.49&0.71&0.83&1.66&\textbf{0.10}&0.37&\textbf{1.14}&1.03&\textbf{0.14}&1.21\\
    ~~~~~0.2&0.0&6.97&10.92&2.31&0.91&0.80&1.44&\textbf{0.11}&0.37&1.56&0.65&\textbf{0.11}&\textbf{0.75}\\
    ~~~~~0.4&0.0&7.21&10.82&2.21&1.26&0.65&1.10&\textbf{0.21}&0.23&2.33&0.79&\textbf{0.08}&\textbf{0.62}\\
    ~~~~~0.6&0.0&7.90&10.68&2.06&0.83&0.75&0.70&\textbf{0.66}&0.10&2.90&0.82&\textbf{0.08}&\textbf{0.63}\\
    ~~~~~0.8&0.0&9.38&10.55&2.14&2.37&1.05&0.79&1.86&0.08&3.02&\textbf{1.26}&\textbf{0.07}&\textbf{0.73}\\
    ~~~~~0.6&-0.8&9.90&10.97&3.17&1.13&0.79&1.16&0.99&0.07&3.59&\textbf{0.75}&\textbf{0.06}&\textbf{0.87}\\
    ~~~~~0.6&-0.4&8.58&10.75&2.53&1.50&0.65&0.80&0.91&1.10&3.13&\textbf{0.67}&\textbf{0.05}&\textbf{0.66}\\
    ~~~~~0.6&0.4&7.72&10.72&1.64&1.97&0.83&1.03&0.43&0.90&2.77&\textbf{0.65}&\textbf{0.10}&\textbf{0.58}\\
    ~~~~~0.6&0.8&8.43&10.86&1.32&2.05&0.94&1.17&\textbf{0.31}&5.10&2.22&0.90&\textbf{0.13}&\textbf{0.67}\\
    \bottomrule
    \end{tabular}}
    \footnotesize{$^{\rm a}$ Arbitrage-free smoothing (AFS) refers to the method proposed in \cite{fengler2009arbitrage}.}
\label{table: out-of-sample errors}
\end{minipage}
\end{center}
\end{table}
To corroborate the assumption, we tabulated the prediction errors in Eqs. \eqref{eq: E_MSE} and \eqref{eq: sigma_MSE}, which signify the predicted option premiums and their IV on the mesh grid for out-of-sample data, in contrast with the ideal values calculated by Eq. \eqref{eq: SABR formula} in the SABR model. These errors are summarized in Table \ref{table: out-of-sample errors}. It is clear that the cubic spline method displays the highest errors in both metrics in all cases. In contrast, MLP and DCNN perform similarly with regard to the premium, while DCNN performs better for volatility. This variation in MLP and DCNN performance across different metrics suggests that DCNN is more efficient in recognizing features on the IV surface, likely due to its integration of no-arbitrage constraints.

\begin{table}
\begin{center}
    \caption{In-sample errors in accuracy and penalty losses. \textbf{Bold values} indicate lower (better) values among the models.}
    {\small\begin{tabular}{@{}cccccccccc}
    \toprule
    \multicolumn{2}{c}{\text {Conditions}} & \multicolumn{6}{c}{\text {In-sample Errors}} \\
    \multicolumn{2}{c}{\text {SABR param.}} & \multicolumn{2}{c}{\text {Cubic Spline}} & \multicolumn{2}{c}{\text {MLP}} & \multicolumn{2}{c}{\text {AFS}$^{\rm a}$} &\multicolumn{2}{c}{\text {DCNN}} \\
    ~~~~~\multirow{2}{*}{$\nu$} & \multirow{2}{*}{$\rho$} & \vspace{3pt} $E_{\text{{\scriptsize MSE}}}$ & $E_\mathcal{P}$ & $E_{\text{{\scriptsize MSE}}}$ & $E_\mathcal{P}$ & $E_{\text{{\scriptsize MSE}}}$ & $E_\mathcal{P}$ & $E_{\text{{\scriptsize MSE}}}$ & $E_\mathcal{P}$ \\
     & & {\scriptsize $\times 10^{-4}$} & {\scriptsize $\times 10^{-3}$} & {\scriptsize $\times 10^{-4}$} & {\scriptsize $\times 10^{-3}$} & {\scriptsize $\times 10^{-4}$} & {\scriptsize $\times 10^{-3}$} & {\scriptsize $\times 10^{-4}$} & {\scriptsize $\times 10^{-3}$} \\
    \midrule 
    ~~~~~0.0 & ~0.0 & 8.44 & 415.0 & 0.47 & 0.38 & $\textbf{0.07}$ & 0.60 & 1.85 & $\textbf{0.13}$ \\
    ~~~~~0.2 & ~0.0 & 8.44 & 414.8 & 0.39 & 0.32 & $\textbf{0.07}$ & 0.68 & 1.64 & $\textbf{0.12}$ \\
    ~~~~~0.4 & ~0.0 & 8.45 & 414.1 & 0.46 & 0.34 & $\textbf{0.16}$ & 0.39 & 1.65 & $\textbf{0.11}$ \\
    ~~~~~0.6 & ~0.0 & 8.48 & 412.6 & 0.45 & 0.36 & $\textbf{0.06}$ & 0.39 & 1.57 & $\textbf{0.11}$ \\
    ~~~~~0.8 & ~0.0 & 8.55 & 410.2 & $\textbf{0.46}$ & 0.35 & 1.63 & 1.70 & 1.48 & $\textbf{0.10}$ \\
    ~~~~~0.6 & -0.8 & 8.59 & 407.9 & $\textbf{0.45}$ & 0.34 & 0.85 & 1.50 & 1.42 & $\textbf{0.09}$ \\
    ~~~~~0.6 & -0.4 & 8.59 & 406.6 & $\textbf{0.43}$ & 0.33 & 0.80 & 1.40 & 1.36 & $\textbf{0.09}$ \\
    ~~~~~0.6 & ~0.4 & 8.60 & 408.4 & 0.43 & 0.33 & $\textbf{0.36}$ & 1.20 & 1.38 & $\textbf{0.09}$ \\
    ~~~~~0.6 & ~0.8 & 8.65 & 411.3 & 0.46 & 0.34 & $\textbf{0.27}$ & 1.50 & 1.46 & $\textbf{0.10}$ \\
    \midrule
    \multicolumn{2}{c}{\text {Quotes 10Jul23}} & 5.75 & 125.9 & $\textbf{0.30}$ & 0.67 & $-^{\rm b}$ & $-^{\rm b}$ & 3.99 & $\textbf{0.34}$ \\
    \multicolumn{2}{c}{\text {Quotes 11Jul23}} & 5.67 & 114.2 & $\textbf{0.48}$ & 0.64 & $-^{\rm b}$ & $-^{\rm b}$ & 3.92 & $\textbf{0.34}$ \\
    \multicolumn{2}{c}{\text {Quotes 12Jul23}} & 5.52 & 119.4 & $\textbf{0.50}$ & 0.90 & $-^{\rm b}$ & $-^{\rm b}$ & 3.91 & $\textbf{0.33}$ \\
    \multicolumn{2}{c}{\text {Quotes 13Jul23}} & 6.01 & 150.2 & $\textbf{0.35}$ & 0.78 & $-^{\rm b}$ & $-^{\rm b}$ & 3.86 & $\textbf{0.34}$ \\
    \multicolumn{2}{c}{\text {Quotes 14Jul23}} & 6.05 & 166.6 & $\textbf{1.46}$ & 1.15 & $-^{\rm b}$ & $-^{\rm b}$ & 5.38 & $\textbf{0.35}$ \\
    \bottomrule
    \end{tabular}}\\
    \footnotesize{$^{\rm a}$ Arbitrage-free smoothing (AFS) refers to the method proposed in \cite{fengler2009arbitrage}.\\
    $^{\rm b}$ AFS requires a rectangular mesh grid, although empirical data has scattered grids.}
\label{table: in-sample errors}
\end{center}
\end{table}
Next, in Table \ref{table: in-sample errors}, we also analyzed the errors in Eqs. \eqref{eq: E_MSE} and \eqref{eq: E_p}, representing calibration performance of option premiums and derivatives implied by the no-arbitrage conditions. While MLP offered the best results in the premium metric, DCNN excelled in the risk metric. This difference in the performance of MLP and DCNN across different metrics substantiates the idea that DCNN's superior capability for identifying features on the IV surface is likely due to its integration of no-arbitrage constraints during learning.

\begin{figure}[H]
    \centering
    \includegraphics[width=0.9\columnwidth]{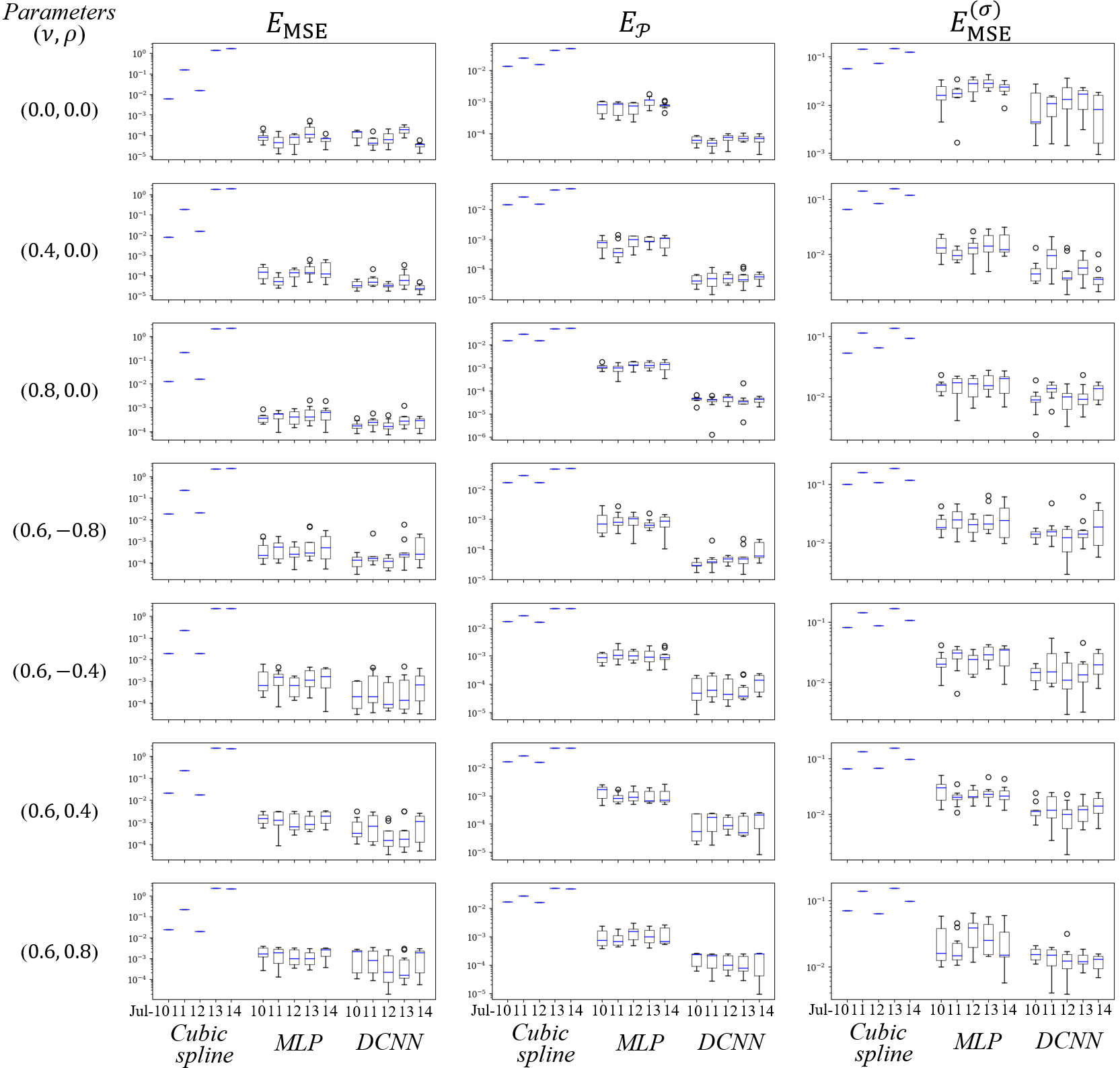}
    \caption{Premium, derivatives penalty and volatility errors with boxplots in out-of-sample of synthetic option premiums on the sparse (and uneven) grids aligned with real market data from 10th to 14th on Jul 2023.}
    \label{Figure: NoArbSS_errors_marketgrids_outofsamples_boxplot}
\end{figure}
\begin{figure}[H]
    \centering
    \includegraphics[width=0.9\columnwidth]{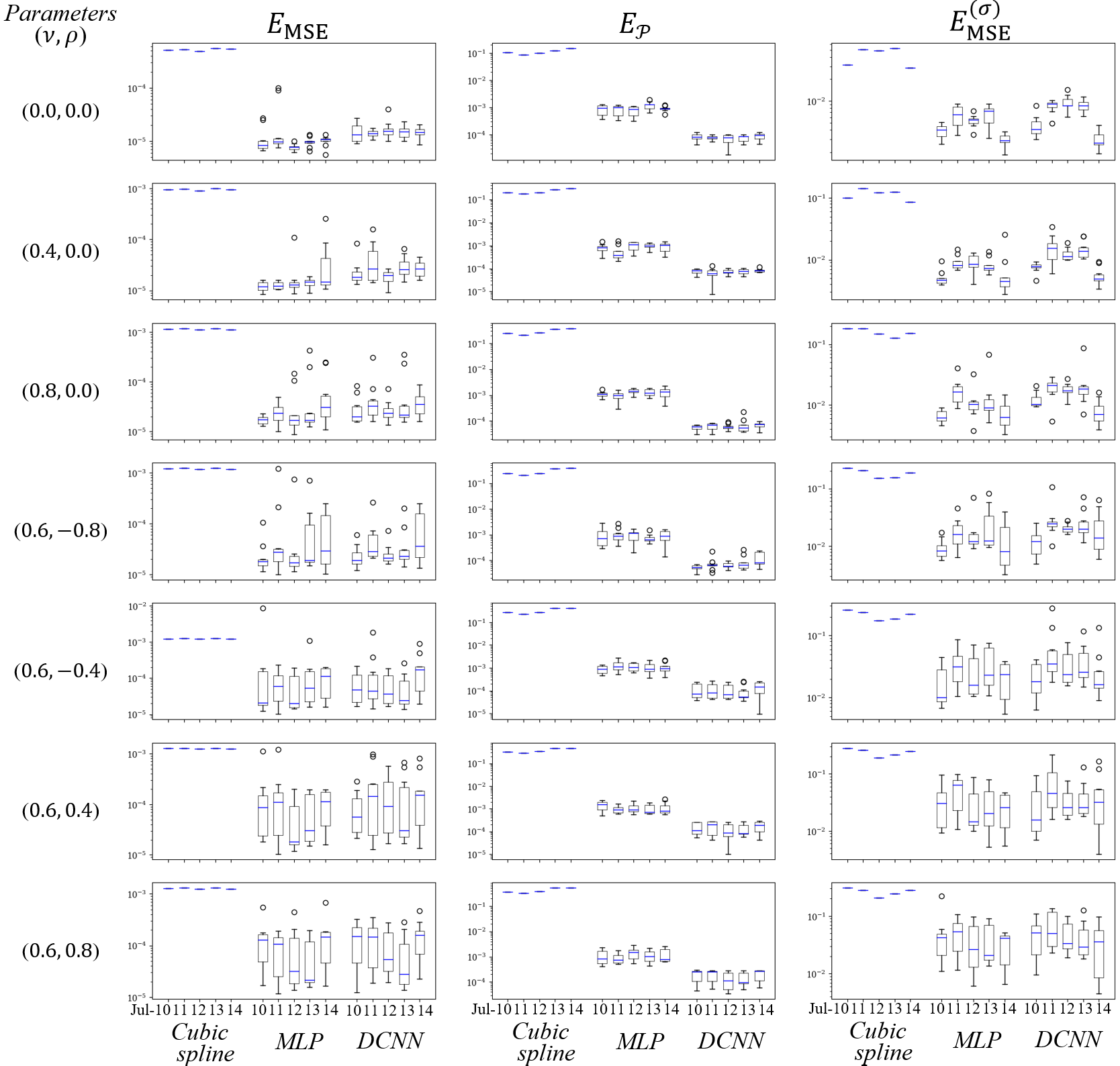}
    \caption{Premium, derivatives penalty and volatility errors with boxplots in in-sample of synthetic option premiums on the sparse (and uneven) grids aligned with real market data from 10th to 14th on Jul 2023.}
    \label{Figure: NoArbSS_errors_marketgrids_insamples_boxplot}
\end{figure}

To evaluate the capability in a more realistic market situation, the following experiment in Figures \ref{Figure: NoArbSS_errors_marketgrids_outofsamples_boxplot} and \ref{Figure: NoArbSS_errors_marketgrids_insamples_boxplot} utilized sparse (and uneven) grid data referring to actual historical traded grids from five days and generated synthetic option premiums on the grids as in-sample training data. In Figure \ref{Figure: NoArbSS_errors_marketgrids_outofsamples_boxplot} tabulated the prediction errors same as that of Table \ref{table: out-of-sample errors}, which signify the predicted option premiums and their IV on the sparse (and uneven) grids for out-of-sample data with dense grids, in contrast with the ideal values in the SABR model. In all test cases, DCNN again showed the best performance with the lowest errors for derivative penalty and implied volatilities across changes in smile and skewness parameters.

In Figure \ref{Figure: NoArbSS_errors_marketgrids_insamples_boxplot}, Training on these sparse, in-sample premiums resulted in higher losses compared to training on even mesh grids. Among the models, DCNN showed the lowest derivative penalties for all test cases, while MLP was superior in terms of accuracy for premium and volatility predictions, different from out-of-sample results.

These results also suggest that DCNN have a greater capability for interpolating premiums from sparse, real-world trading data compared to the other methods. DCNN provided the most stable results for implied volatility, indicating it can effectively generate a surface capturing the smile and skewness features on the predictions. This demonstrates the DCNN's ability to characterize the implied volatility surface more precisely.

\begin{figure}[htbp]
    \centering
    \includegraphics[width=1.0\columnwidth]{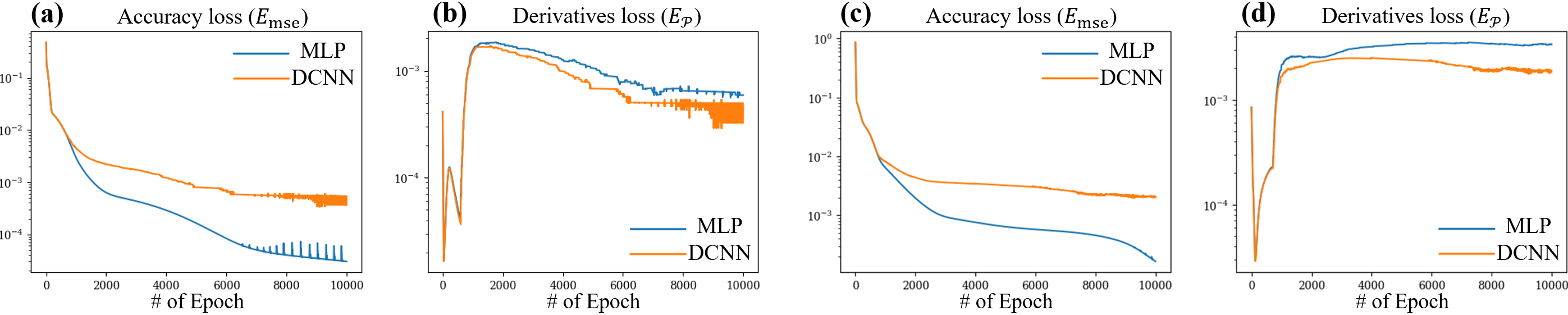}
    \caption{The accuracy (MSE) and loss due to the derivatives in the experiments on synthetic data  \textbf{(a-b)} and real market data \textbf{(c-d)} with a logarithmic scale by learning epochs.}
    \label{Figure: NoArbSS_learning_results}
\end{figure}
Furthermore, Figure \ref{Figure: NoArbSS_learning_results} depicts the MSE loss and the penalty loss due to derivative terms across various epochs. A trade-off between solution accuracy and the penalty from derivative terms is visible throughout the learning process. In Figure \ref{Figure: NoArbSS_learning_results}\textbf{(b)} and \textbf{(d)}, a decrease in penalty related to the derivative term is seen for DCNNs. This trend shows a trade-off between accuracy and penalty from derivative terms, brought about by the effort to comply with no-arbitrage constraints during learning.

\begin{figure}[htbp]
    \centering
    \includegraphics[width=\columnwidth]{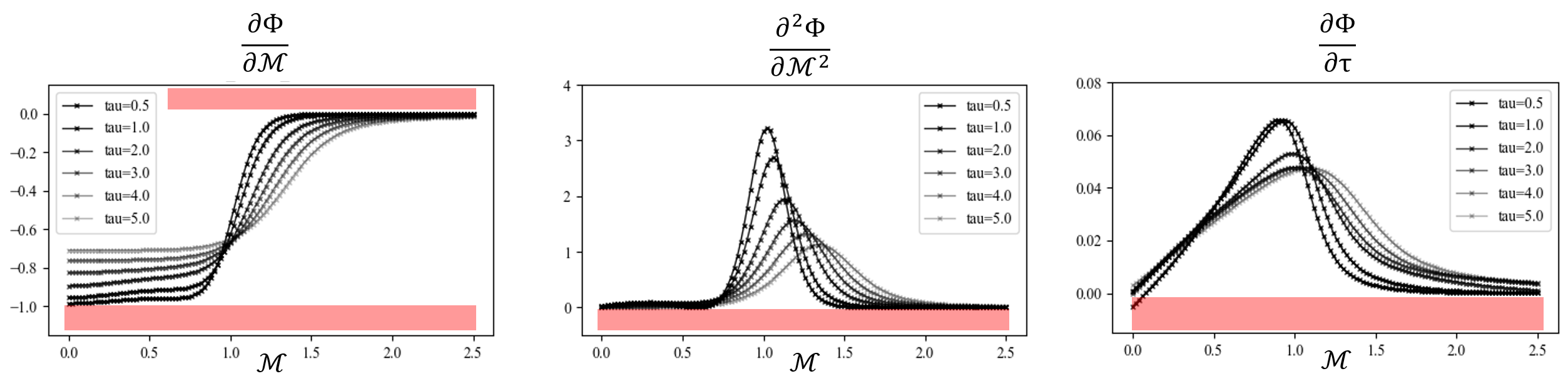}
    \caption{Derivative profiles by DCNN. It shows the first and second differential values of $\mathcal{M}$ (left and centre) and the first differential values of $\tau$ (right), and each line with sliced in $\tau \in [0.5,1.0,2.0,3.0,5.0]$. The colored area indicates breaking the no-arbitrage condition aligned with each inequality in the no-arbitrage constraints in Eq.~\eqref{eq: no-arbitrage conditions}.}
    \label{Figure: NoArbSS_result_risk_profiles}
\end{figure}
In order to know the characteristics of the function surface in detail, Figure \ref{Figure: NoArbSS_result_risk_profiles} shows the first and second differential values of $\mathcal{M}$ (up and center) and the first differential values of $\tau$ of the surface in each model surface sliced at $\tau\in[0.5,1.0,2.0,3.0,5.0]$. In financial terms, each derivative corresponds to risk parameters in orders: \textit{Dual Delta}, \textit{Dual Gamma}, and \textit{Dual Theta} as described in \cite{wystup2002vanilla}. We can observe that DCNN successfully prevents the results from breaking conditions for the derivative values, as represented by the shaded areas in Figure \ref{Figure: NoArbSS_result_risk_profiles}. Note that DCNN is still a soft constraint approach and does not secure fully no-arbitrage conditions.

\begin{figure}[htbp]
    \centering
    \includegraphics[width=0.95\columnwidth]{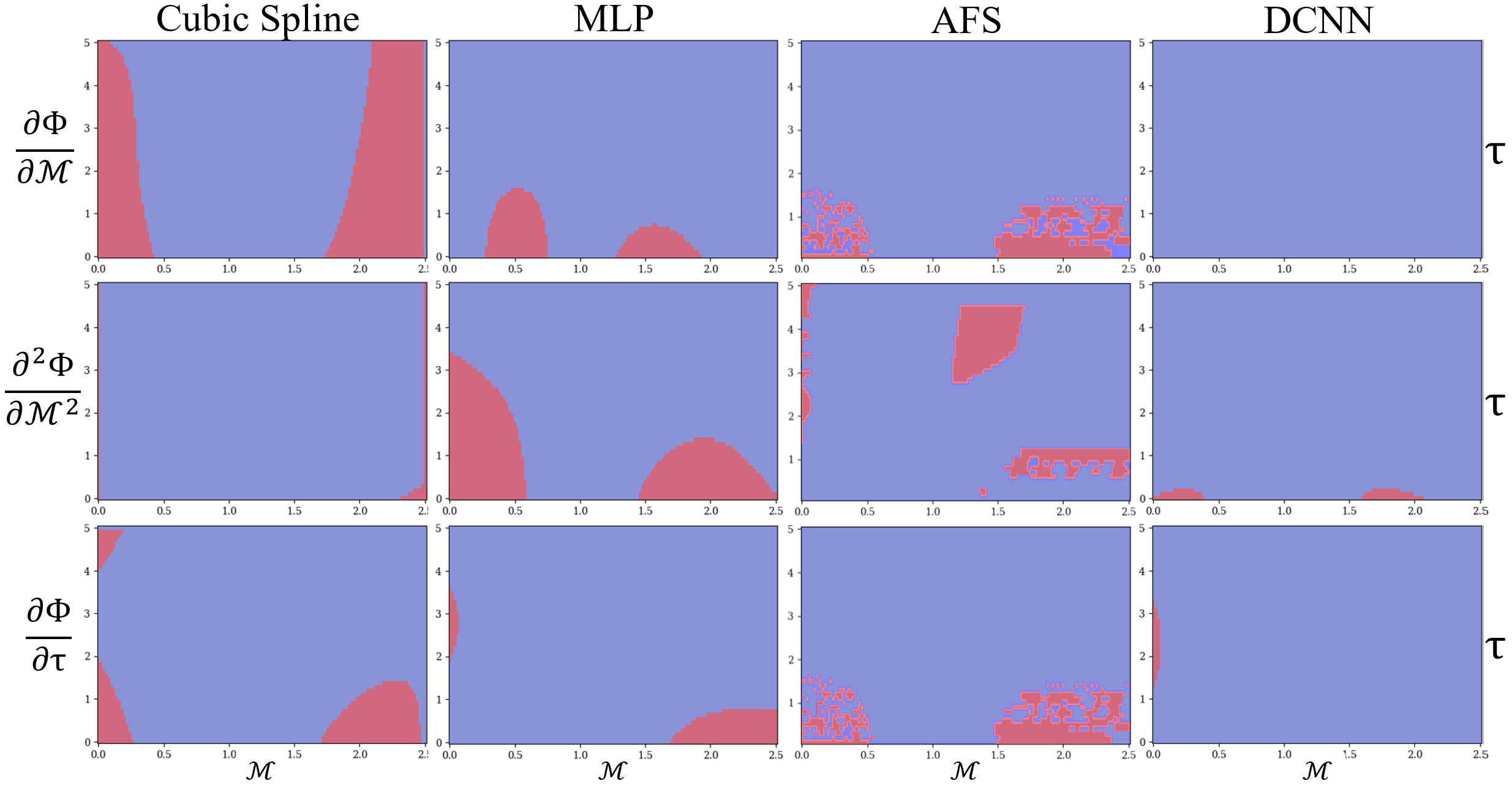}
    \caption{Risk profiles (i.e. derivative terms) comparison by option premium surface. The red-coloured area indicates breaking no-arbitrage conditions.}
    \label{Figure: NoArbSS_arbitrary_results}
\end{figure}
Lastly, Figure \ref{Figure: NoArbSS_arbitrary_results} shows the first and second differentials of predicted values in $\mathcal{M}$ and $\tau$ in Eq. \eqref{eq: no-arbitrage conditions}. DCNN effectively ensures that the results do not infringe on the conditions for derivative values, each of which aligns with an inequality constraint in Eq. \eqref{eq: no-arbitrage conditions}. In Figure \ref{Figure: NoArbSS_arbitrary_results}, we see that the plots for DCNN have fewer areas in red, denoting the regions where the no-arbitrage conditions on the derivatives are broken. Consequently, these findings prove that DCNN is a powerful tool that aids practitioners not only in understanding market volatility dynamics but also in managing risks effectively.

\section{Discussion}
The specific contributions of this paper are as follows: 
\begin{enumerate}
\setlength{\leftskip}{-10pt}
\item The development of a new calibration framework for option premiums: This study utilized the DCNN algorithm, powered by deep learning, which efficiently computes derivatives with automatic differentiation and introduces differential constraints, allowing convergence to the appropriate function surface with accurate detection of volatility characteristics whilst adhering to no-arbitrage conditions.
\item Effectiveness of DCNN in capturing the features of the IV surface: The study utilized the SABR model to assess the ability of DCNN to represent the capability of smile and skewness features in IV surfaces. DCNN significantly improved the interpolation of the premium surface for volatility smile and skewness due to deep learning with no-arbitrage constraints.
\item Trade-offs between price accuracy and derivative penalty: The findings demonstrate that as the learning process proceeds, there's a decrease in the penalty associated with the derivative term, reflecting a balance between accuracy and penalty from derivative terms. The results emphasize the network's efforts to comply with no-arbitrage constraints during learning, enhancing the model's overall effectiveness and robustness.
\end{enumerate}
\paragraph{Limitations} DCNNs have inherent methodological limitations rooted in their fundamental definitions. Specifically, DCNNs are designed for simple MLPs, which are fully connected feedforward networks based on the universal approximation theorem. DCNNs have not yet been implemented for other network architectures where this theorem's security is not fully guaranteed. Additionally, the main text was unable to definitively prescribe weightings for each loss term in multi-task deep learning contexts. However, we offer one solution using a soft-weighting approach for self-adaptive DCNNs, detailed in Appendix \ref{appendix: Self-Adaptive Derivative-Constrained Neural Network}.

\section{Conclusion}
This paper presents Derivative-Constrained Neural Networks (DCNN), a neural network algorithm for approximating solutions to partial differential equations. It employs derivative values obtained through automatic differentiation, enhancing the overall accuracy and interpretability of the solution. This includes meeting constraints such as no-arbitrage, capturing volatility smile and skewness effects, and addressing the limitations of traditional calibration methods in scenarios with sparse training data. In our experimental setup, DCNN shows better results in identifying features 
 of smile and skewness on the implied volatility surface, likely due to its incorporation of no-arbitrage constraints during the learning process. This study uses the SABR Stochastic Volatility model to demonstrate the improved interpolation of the premium surface and risk profiles. The findings on both simulated and real data highlight the potential of DCNN as a tool for understanding market dynamics and managing risk. It provides a data-driven solution to calibration problems, allowing for an accurate representation of both option premium and volatility surfaces.

\section*{Acknowledgement}
The authors are thankful to the Department of Computer Science, University College London, for providing us with the resources to perform this case study.


\section*{Supplemental material}

The conceptual codes for this study are available for peer review at a private link. The codes would be made publicly available on GitHub prior to the publication of the paper. The datasets generated and/or analyzed during the current study are not publicly available due to privacy and data security considerations but are available from the corresponding author upon reasonable request.

\bibliography{references}

\appendices
\section{The first and second derivatives of  Multi-layer Perceptron (MLP)}
\label{appendix: The first and second derivatives of MLP}
This section extends the work on the first and second derivatives of MLP to cross derivatives. In second derivative calculations, Eq.~\eqref{eq: nabla squared layer} can be applied for the selections of second-order partial derivatives, but not for the mixed partial derivatives. The formulations in this section demonstrate how cross-derivatives can be incorporated to extend derivative-constrained neural networks (DCNNs).

Following the notation in Section \ref{section: Derivative-Constrained Neural Network}, we can redefine $\nabla^2$ in Eq.~\eqref{eq: definition of nabla and nabla squeared} with cross derivatives using the Hessian matrix $\mathbf{H}$,
\begin{equation}
\nabla^2\Phi_{W,b}\left(\mathbf{x}\right):=\mathbf{H}_\Phi=\left[\begin{array}{cccc}
\pdv[2]{\Phi}{x_1} & \pdv[2]{\Phi}{x_1}{x_2} & \cdots & \pdv[2]{\Phi}{x_1}{x_n} \\
\pdv[2]{\Phi}{x_2}{x_1} & \pdv[2]{\Phi}{x_2} & \cdots & \pdv[2]{\Phi}{x_2}{x_n} \\
\vdots & \ddots & \ddots & \vdots \\
\pdv[2]{\Phi}{x_n}{x_1} & \pdv[2]{\Phi}{x_n}{x_2} & \cdots & \pdv[2]{\Phi}{x_n} \\
\end{array}\right].
\label{eq: nabla squared Hessian ver}
\end{equation}
Here, the entry of the $i$-th row and the $j$-th column is
\begin{equation}
    \left(\mathbf{H}_f\right)_{i j}=\frac{\partial^2 f}{\partial x_i \partial x_j}.
\end{equation}
Similarity, we reconsider the second derivative of each layer $\nabla^2_l \in \mathbb{R}^{d_l \times n \times n}$ at the $l$-th layer defined in Eq.~\eqref{eq: nabla squared layer} with cross derivatives, 
\begin{equation}
    \nabla^2_l 
    := \pdv[2]{\mathbf{x}_{l}}{\mathbf{x}} = \left[ \mathbf{H}_{x^{(l)}_1}, \ldots, \mathbf{H}_{x^{(l)}_{d_l}} \right]
\end{equation}
where, the elements of the Hessian matrix $\mathbf{H}_{x^{(l)}_i}$ are
\begin{equation}
\mathbf{H}_{x^{(l)}_i}=\left[\begin{array}{ccc}
\pdv[2]{x^{(l)}_i}{x_1} & \cdots & \pdv[2]{x^{(l)}_i}{x_1}{x_n} \\
\vdots & \ddots & \vdots \\
\pdv[2]{x^{(l)}_i}{x_n}{x_1} & \cdots & \pdv[2]{x^{(l)}_i}{x_n} \\
\end{array}\right].
\end{equation}

Calculating the derivatives of tensor equations, a process known as tensor calculus, is crucial in machine learning. One significant aspect to consider is the efficiency of evaluating these equations and their derivatives, which depends on the way these expressions are represented. Here, Eqs.~\eqref{eq: nabla layer} and \eqref{eq: nabla squared layer} - which contain cross derivatives and are characterized by tensor computations in an automatic differentiation framework.

\paragraph{\textbf{Tensor calculation}} In tensor calculus \cite{laue2020simple}, for tensors $A, B$, and $C$ the multiplication of $A$ and $B$ can be written as
\begin{equation}
    C\left[s_3\right]=\sum_{\left(s_1 \cup s_2\right) \backslash s_3} A\left[s_1\right] \cdot B\left[s_2\right]=A \ast_{\left(s_1, s_2, s_3\right)} B
\end{equation}
where $C$ is the result tensor and $s_1, s_2$, and $s_3$ are the index sets of the left argument, the right argument, and the result tensor, respectively. The summation index is excluded from the index set of the result tensor $s_3 \subseteq\left(s_1 \cup s_2\right)$ explicitly represents the index set of $C$, which is always a subset of the union of $s_1$ and $s_2$.

\vspace{7pt}
Based on the definition provided, tensor multiplication can be described succinctly with fewer summation symbols. Furthermore, this notation closely resembles the tensor multiplication \texttt{einsum} found in Python packages. As an illustration, the inner product of matrices $A$ and $B$ can be expressed as

\begin{equation}
    A*_{\left(ik, kj, ij\right)}B = \sum_k A_{i k} \cdot B_{k j}= A \cdot B.
\end{equation}

Subsequently, we present a formulation for the first and second derivatives of the function $\Phi$. Note that the product chain rule can be employed between the $l$-th and $(l-1)$-th layers, given that each layer is fully connected in the feedforward network.

Initially, we examine the first-order derivatives of each layer's output $\mathbf{x}{l}$ with respect to the input features $\mathbf{x}$. Note that the input features are $\mathbf{x} \in \mathbb{R}^n$ and $\mathbf{x} = x_1,\ldots,x_n$. The outputs of the $l$-th layer are denoted as $\mathbf{x}^{(l)} \in \mathbb{R}^{d_l}$, where $\mathbf{x}^{(l)} = x_1^{(l)},\ldots,x_{d_l}^{(l)}$. Consequently, each element of the first derivative at the first layer is formulated as
\begin{equation}
    \pdv[]{x_{j}^{\left(1\right)}}{x_i} = (f_1^\prime \circ A_1 \left(\mathbf{x}\right))_j \times (W_{1}^\top)_{ji}.
\end{equation}
By employing tensor calculation, we can succinctly represent $\left(\nabla_1\right)_{ji} = \pdv[]{x_{j}^{\left(1\right)}}{x_i}$ as
\begin{equation}
\nabla_1 = f_1^\prime \circ A_1 \left(\mathbf{x}\right) \ast_{\left(j, ji, ji\right)} W_{1}^\top,
\end{equation}

For $l=2\ldots,L-1$, we adhere to the linear algebra notations found in \cite{laue2020simple}, and rewrite Eq.~\eqref{eq: nabla layer} using tensor notation with the chain product rule
\begin{equation}
\nabla_l 
= \left\{ \left(f_l^\prime \circ A_l \left(\mathbf{x}_{l-1}\right)\right) \ast_{\left(j, ji, ji\right)} W_{l}^\top \right\}\nabla_{l-1}.
\label{eq: nabla_1 Einstein}
\end{equation}
$\nabla_L$ is expressed because the last layer lacks an activation function, 
\begin{equation}
\nabla_L = \left\{ A_L \left(\mathbf{x}_{L-1}\right) \ast_{\left(j, ji, ji \right)} W_{L}^\top \right\}\nabla_{L-1}.
\end{equation}
Correspondingly, leveraging the above equations, we can formulate the second derivatives with cross derivatives. At this juncture, each element of the second derivative at the first layer is formulated as
\begin{equation}
 \mathbf{H}_{x^{(1)}_j}
    = \left(f_1^{\prime\prime}\circ A_1\left(\mathbf{x}\right) \ast_{\left(j, ji, ji\right)} \left(W_{1}^\top\right)\right)_j \otimes \left(W_{1}^\top \right)_j,
\end{equation}
where $\otimes$ indicates the dyadic product of vectors. Utilizing the tensor calculation, we can represent $\left(\nabla^2_1\right)_{j} = \mathbf{H}_{x^{(1)}_j}$ as
\begin{equation}
\nabla^2_1 
    = \left\{f_1^{\prime\prime}\circ A_1\left(\mathbf{x}\right) \ast_{\left(j, ji, ji\right)} \left(W_{1}^\top\right)\right\} \ast_{\left(ji, jk, jik\right)} \left(W_{1}^\top \right).
\end{equation}
When $l=2\ldots,L-1$, the partial derivatives of the outputs of the $l$-th layer with respect to $\mathbf{x}$ is calculated using the product rule, i.e.  $(u \cdot v)^{\prime}=u^{\prime} \cdot v+u \cdot v^{\prime}$, for Eq.~\eqref{eq: nabla_1 Einstein} with $\nabla_{l-1}$ and $\nabla^2_{l-1}$ as
\begin{equation}
\begin{split}
    \nabla^2_l 
    = & \left\{f_l^{\prime\prime}\circ A_l\left(\mathbf{x}_{l-1}\right) \ast_{\left(j, ji, ji\right)} \left(W_{l}^\top \nabla_{l-1}\right)\right\} \ast_{\left(ji, jk, jik\right)} \left(W_{l}^\top \nabla_{l-1}\right) \\
    & +\left\{ f_l^\prime \circ A_l \left(\mathbf{x}_{l-1}\right) \ast_{\left(j, ji, ji\right)} W_{l}^\top \right\} \ast_{\left(jm, mik, jik\right)} \nabla^2_{l-1},   
\end{split}
\end{equation}
where $f_l^{\prime}$ and $f_l^{\prime \prime}$ are the first and second-order derivatives of the activation function at the $l$-th layer, respectively. Lastly, the second derivative for the last layer, which is equivalent to $\nabla^2 \Phi\left( \mathbf{x} \right)$, is obtained as 
\begin{equation}
    \nabla^2 \Phi\left( \mathbf{x} \right) 
    =\nabla^2_L 
    = \left\{ A_L \left(\mathbf{x}_{L-1}\right) \ast_{\left(j, ji, ji\right)} W_{L}^\top \right\} \ast_{\left(jm, mik, jik\right)} \nabla^2_{L-1}.
    \label{eq: nabla_squared Einstein}
\end{equation}

The second-order derivatives with cross derivatives of an MLP, as delineated in Eq.~\eqref{eq: nabla_squared Einstein}, can be seamlessly integrated into Algorithm 1. As a natural extension of DCNNs for financial applications, additional cross-derivative sensitivities such as Vanna, Charm, and cross-Gamma could be incorporated for multi-asset products. 

\section{The selection of Network configulations}
\label{appendix: The selection of Network setting}

\subsection{Differentiability of activation functions}
\label{appendix: Differentiability of activation functions}
This section summarizes the network's requirement of the introduced DCNN and compares network configurations to performance for guiding principles for selecting appropriate configurations for its models and applications.

First of all, consideration of the differentiability of the activation function is essential in selecting the activation function of the DCNN. To adapt DCNN, the function of the network requires at least a second differentiable because The derivative formula of the activation function's derivatives, $f^\prime$, and $f^{\prime\prime}$, are required.

\begin{table}[htbp]
\small
\centering
\begin{threeparttable}[hbtp]
    \renewcommand{\arraystretch}{1.5}
    \caption{Derivatives functions of the typical activation function. PReLU in \cite{he2015delving} and MPELU in \cite{li2018improving} are generalized and unified for ReLU and ELU.}
    \label{table:activation_function_derivatives}
    \centering
        \begin{tabular}{cccc}
        \hline Name & Activation function($f$) & First derivatives($f^{\prime}$) & Second derivatives($f^{\prime\prime}$) \\
        \hline
        Sigmoid 
            & {\tiny ${\left(1+e^{-x}\right)}^{-1}$} 
                & {\tiny ${e^{-x}}{(1+e^{-x})^{-2}}$} 
                    & {\tiny ${e^{-2 x}\left(1-e^x \right)}{\left(1+e^{-x}\right)}^{-3}$} \\
        Softplus
            & {\tiny $\ln \left(e^x + 1\right)$ }
                & {\tiny ${\left(1+e^{-x}\right)}^{-1}$}
                    & {\tiny ${e^{-x}}{(1+e^{-x})^{-2}}$} \\
        Hyperbolic Tangent 
            & {\tiny ${\left(e^{x}-e^{-x}\right)}{\left(e^{x}+e^{-x}\right)}^{-1}$} 
                & {\tiny ${4}{\left(e^x+e^{-x}\right)^{-2}}$} 
                    & {\tiny $-8{\left(e^x-e^{-x}\right)}{\left(e^x+e^{-x}\right)^{-3}}$} \\
        ReLU (PReLU) 
            & {\tiny $\begin{cases} x &\text{if}\ \ x > 0 
            \\ 0 &\text{if}\ \ x = 0 
            \\ ax &\text{if}\ \ x < 0 \end{cases}$}
                & {\tiny $\begin{cases} 1 &\text{if}\ \ x > 0 
            \\ \text{not defined} &\text{if}\ \ x = 0
            \\ a &\text{if}\ \ x < 0 \end{cases}$} 
                    & {\tiny $\begin{cases} 0 &\text{if}\ \ x \neq 0 
            \\ \text{not defined} &\text{if}\ \ x = 0 \end{cases}$} \\
        ELU (MPELU) 
            & {\tiny $\begin{cases} x & \text{if}\ \ x > 0\\ 0 &\text{if}\ \ x = 0 \\ \alpha(e^{\beta x}-1) & \text{if}\ \ x < 0 \end{cases}$} 
                & {\tiny $\begin{cases} 1 & \text{if}\ \ x > 0 \\ \text{not defined} &\text{if}\ \ x = 0
            \\ \alpha \beta e^{\beta x} & \text{if}\ \ x < 0 \end{cases}$} 
                    & {\tiny $\begin{cases} 0 & \text{if}\ \ x > 0 \\ \text{not defined} &\text{if}\ \ x = 0\\ \alpha \beta^2 e^{\beta x} & \text{if}\ \ x < 0 \end{cases}$} \\
        \hline
        \end{tabular}
\end{threeparttable}
\end{table}

In addition to that, C2 continuity is usually required for almost all problems by DCNN because of the requirement of continuity of the first and second derivative surfaces. In Table \ref{table:activation_function_derivatives}, The derivatives of the Rectified Linear Unit (ReLU) and Exponential Linear Unit (ELU) functions are undefined at $x=0$. These are typically set to constant values to prevent errors during the backpropagation process in the vanilla process, which doesn't use DCNN approach. In Table \ref{table:activation_function_derivatives_y}, the high-order derivatives with the output of MLP ($y$) are enumerated for typical activation functions. It is noted that the backpropagation with DCNN requires additional settings, which is a higher-order differentiable activation function by reverse-mode of automatic differentiation as shown in the third derivatives in Table \ref{table:activation_function_derivatives_y}.

\begin{table}[htbp]
\small
\centering
\begin{threeparttable}[hbtp]
    \renewcommand{\arraystretch}{1.5}
    \caption{Derivatives functions of the output in the typical activation function. The output is defined as $y= f(x)$ in Table \ref{table:activation_function_derivatives}. PReLU\cite{he2015delving} and MPELU\cite{li2018improving} are generalized and unified for well-known ReLU and ELU.}
    \label{table:activation_function_derivatives_y}
    \centering
        \begin{tabular}{cccc}
        \hline Name & First derivatives($f^{\prime}$) & Second derivatives($f^{\prime\prime}$) & Third derivatives \\
        \hline
        Sigmoid 
            & {\tiny $y\left(1-y\right)$} 
                & {\tiny $y\left(1-y\right)\left(1-2y\right)$}
                    & {\tiny $y\left(1-y\right)\left(1-6y+6y^2\right)$}\\
        Softplus
            & {\tiny $1-e^{-y}$}
                & {\tiny ${e^{-y}}{(1-e^{-y})}$}
                        & {\tiny ${e^{-2y}}{(1-e^{-y})}{(2-e^{y})}$}\\
        Hyperbolic Tangent 
            & {\tiny $1-y^2$} 
                & {\tiny $-2y\left(1-y^2\right)$}
                    & {\tiny $\left(1-y^2\right)\left(6y^2-2\right)$} \\
        ReLU (PReLU) 
            & {\tiny $\begin{cases} 1&\text{if}\ y > 0\\ \text{not defined} &\text{if}\ \ y = 0\\ a&\text{if}\ y < 0 \end{cases}$} 
                    & {\tiny $\begin{cases} 0 &\text{if}\ \ y \neq 0 
            \\ \text{not defined} &\text{if}\ \ y = 0 \end{cases}$} 
                        & {\tiny $\begin{cases} 0 &\text{if}\ \ y \neq 0 
            \\ \text{not defined} &\text{if}\ \ y = 0 \end{cases}$}\\
        ELU (MPELU) 
            & {\tiny $\begin{cases} 1 & \text{if}\ \ y > 0\\ \text{not defined} &\text{if}\ \ y = 0\\ \beta \left(y+\alpha  \right)& \text{if}\ \ y < 0 \end{cases}$} 
                    & {\tiny $\begin{cases} 0 & \text{if}\ \ y > 0 \\ \text{not defined} &\text{if}\ \ y = 0\\ \beta^2 \left(y+\alpha  \right) & \text{if}\ \ y < 0 \end{cases}$}
                        & {\tiny $\begin{cases} 0 & \text{if}\ \ y > 0 \\ \text{not defined} &\text{if}\ \ y = 0\\ \beta^3 \left(y+\alpha  \right) & \text{if}\ \ y < 0 \end{cases}$} \\
        \hline
        \end{tabular}
\end{threeparttable}
\end{table}

Finally, adapting DCNN to include derivative information necessitates two careful considerations. First, we need to ensure the representational capability of the derivative surface, which implies that the selected activation function should be adept at approximating the shape and range of the derivative surface using its own derivative function. Second, we must confirm the existence of a higher-order derivative of the derivative surfaces for the backpropagation in gradient-based learning. One of our findings suggests that DCNN shows a trade-off in the efficiency and accuracy of the learning, inferring that the efficient changes in performance are reflected by alterations in the property of higher-order derivatives of activation functions.

\subsection{Comparison of network configulations}
The architecture and hyperparameters used to configure neural networks impact their performance on machine learning tasks. Building on the property analysis, this section provides a comparative functional assessment focused on model accuracy, i.e., parameterization impact of networks on predictive performance.

\begin{landscape}
\thispagestyle{empty}
\begin{table}[htbp]
\centering
\caption{Performance comparison of each error by the configurations of neural networks and packages, displayed as mean values with standard deviations. The other settings are similar to that in Section \ref{subsection: Experimental design}$^{\rm a}$.}
{\small
    \centering\begin{tabular}{@{}ccccrlrlrlrlrlrl}
    \toprule
    \multicolumn{4}{c}{Network configuration (\# of)} &  \multicolumn{6}{c}{MLP} & \multicolumn{6}{c}{DCNN}\\
    \multirow{2}{*}{\begin{tabular}{c} Act. func.\\$(f)$\end{tabular}} & \multirow{2}{*}{\begin{tabular}{c}Layers\\$(L)$\end{tabular}}&\multirow{2}{*}{\begin{tabular}{c}Neurons\\$(d_l)$\end{tabular}}&\multirow{2}{*}{\begin{tabular}{c}Param.\\$(W, b)$\end{tabular}}& $E_{\rm \tiny MSE}$ & (std) & $E_{\rm \tiny MSE}^{(\sigma)}$ & (std) & $E_\mathcal{P}$ & (std) & $E_{\rm \tiny MSE}$ & (std) & $E_{\rm \tiny MSE}^{(\sigma)}$ & (std) & $E_\mathcal{P}$ & (std) \\
    &&&& \multicolumn{2}{c}{\scriptsize $\times 10^{-4}$} & \multicolumn{2}{c}{\scriptsize $\times 10^{-2}$} & \multicolumn{2}{c}{\scriptsize $\times 10^{-3}$} & \multicolumn{2}{c}{\scriptsize $\times 10^{-4}$} & \multicolumn{2}{c}{\scriptsize $\times 10^{-2}$} & \multicolumn{2}{c}{\scriptsize $\times 10^{-3}$} \\
    \hline
    elu&3&16&337&0.17&(0.06)&0.43&(0.13)&0.24&(0.14)&0.33&(0.18)&0.37&(0.15)&0.03&(0.01)\\
    elu&5&16&881&0.16&(0.16)&0.38&(0.23)&0.45&(0.15)&0.19&(0.07)&0.37&(0.18)&0.03&(0.01)\\
    elu&9&16&1,969&0.18&(0.13)&0.49&(0.23)&1.64&(0.37)&0.93&(1.35)&0.63&(0.36)&0.04&(0.02)\\
    elu&3&64&4,417&0.25&(0.10)&0.44&(0.09)&0.29&(0.13)&0.37&(0.32)&0.49&(0.31)&0.03&(0.01)\\
    elu&5&64&12,737&18.48&(44.52)&2.68&(4.51)&0.98&(0.82)&4.17&(7.85)&0.85&(0.74)&0.05&(0.03)\\
    elu&3&128&17,025&3.16&(5.58)&1.07&(1.15)&0.34&(0.16)&1.12&(1.30)&0.70&(0.43)&0.06&(0.04)\\
    elu&9&64&29,377&1231.99&(671.56)&51.60&(23.49)&0.46&(0.90)&782.23&(743.31)&34.62&(26.99)&0.51&(0.83)\\
    elu&5&128&50,049&167.39&(460.94)&8.23&(17.56)&0.89&(0.97)&37.20&(62.49)&4.44&(4.93)&0.26&(0.31)\\
    elu&9&128&116,097&1279.89&(484.98)&53.70&(19.79)&24.44&(69.17)&1420.40&(527.32)&59.24&(23.09)&2.06&(6.03)\\
    \cdashline{1-16}softplus&3&16&337&2.04&(3.68)&0.99&(0.72)&0.85&(0.69)&1.45&(2.11)&0.73&(0.44)&0.07&(0.04)\\
    softplus&5&16&881&0.04&(0.01)&0.15&(0.09)&0.84&(0.15)&0.14&(0.09)&0.35&(0.22)&0.02&(0.00)\\
    softplus&9&16&1,969&0.49&(0.54)&0.44&(0.37)&1.35&(0.27)&0.69&(0.48)&0.76&(0.27)&0.03&(0.01)\\
    softplus&3&64&4,417&0.08&(0.01)&0.31&(0.17)&0.24&(0.04)&0.30&(0.39)&0.33&(0.17)&0.03&(0.01)\\
    softplus&5&64&12,737&0.63&(1.75)&0.38&(0.48)&1.46&(1.21)&0.09&(0.04)&0.25&(0.16)&0.02&(0.01)\\
    softplus&3&128&17,025&0.63&(1.14)&0.52&(0.36)&0.16&(0.06)&0.27&(0.08)&0.36&(0.11)&0.02&(0.00)\\
    softplus&9&64&29,377&1240.29&(618.95)&48.65&(23.80)&0.54&(1.09)&1236.91&(618.34)&48.43&(23.92)&0.01&(0.01)\\
    softplus&5&128&50,049&0.63&(1.07)&0.48&(0.40)&2.52&(0.84)&0.42&(0.62)&0.50&(0.39)&0.04&(0.02)\\
    softplus&9&128&116,097&1549.45&(0.75)&60.53&(0.03)&0.00&(0.00)&1548.61&(1.09)&60.49&(0.06)&0.00&(0.00)\\
    \cdashline{1-16}tanh&3&16&337&0.19&(0.22)&0.47&(0.27)&0.82&(0.42)&0.14&(0.09)&0.37&(0.18)&0.03&(0.01)\\
    tanh&5&16&881&0.16&(0.31)&0.47&(0.49)&1.22&(0.21)&0.17&(0.10)&0.54&(0.42)&0.02&(0.01)\\
    tanh&9&16&1,969&0.21&(0.18)&0.48&(0.25)&2.89&(0.69)&1.05&(1.12)&1.05&(0.76)&0.05&(0.03)\\
    tanh&3&64&4,417&1.38&(2.34)&0.85&(0.79)&1.58&(0.46)&0.17&(0.12)&0.52&(0.38)&0.02&(0.01)\\
    tanh&5&64&12,737&161.35&(462.66)&7.57&(17.72)&232.18&(333.81)&174.40&(491.32)&8.21&(18.66)&0.27&(0.26)\\
    tanh&3&128&17,025&175.21&(516.90)&6.51&(15.99)&3.42&(1.75)&4.32&(8.50)&1.07&(0.71)&0.07&(0.05)\\
    tanh&9&64&29,377&1423.25&(246.61)&55.63&(13.42)&74.81&(224.43)&1606.31&(125.70)&62.60&(4.59)&0.00&(0.00)\\
    tanh&5&128&50,049&902.03&(737.56)&35.80&(27.55)&51.63&(119.59)&906.51&(738.74)&36.15&(27.83)&0.20&(0.45)\\
    tanh&9&128&116,097&1701.27&(303.26)&68.96&(16.89)&0.00&(0.00)&1459.35&(138.41)&56.34&(6.72)&0.00&(0.00)\\
  \hline
\end{tabular}}
\footnotesize{$^{\rm a}$ Experimental conditons are $N=575, M=286$ and $10,000$ epochs repeated 10 times in the synthesized data with $\rho=-0.4$ and $\nu=0.6$ in the SABR model. \\ $^{\rm b}$ In the case of the ELU function, $\alpha = \beta = 1$ as applied in Table \ref{table:activation_function_derivatives}.}
\label{table: comp time in the various conditions}
\end{table}
\vfill
\raisebox{-10pt}{\makebox[\linewidth]{\thepage}}
\end{landscape}

Table \ref{table: comp time in the various conditions} compares the predictive performance of MLP and DCNN models under various architectural configurations. Increasing depth and width generally decreases MSE up to a point before overfitting occurs. Softplus activation tends to provide the lowest errors for shallow networks. Tanh produces the lowest errors for medium depth/width ELU activation, resulting in some instability for deeper models. The analysis shows the importance of tuning network configuration and activation functions jointly to optimize performance and stability. Overall, guidelines indicate that 2-4 layers with 16-64 units per layer work well across functions; deeper models require more careful activation selection.

Overfitting manifests in very high volume parameter settings, an inherent challenge posing barriers to generalization. As explored by \cite{ying2019overview}, some regularization methods can give improvements to prevent units from co-adapting too much in such overfitting. One idea of the solution for the overfitting is to randomly drop units (along with their connections) from the neural network during training, introduced in \cite{srivastava2014dropout}.

The contrasts along dimensions of properties and functional performance provide valuable insights to inform appropriate activation function selection for different model architectures and machine learning problem settings. Overall there remain ample opportunities to refine activation choices and improve generalization across network complexities when applying neural networks to financial pricing.

\section{Computational complexity}
\label{appendix: Computational complexity}
To evaluate the stability of the DCNN under various configurations, we conducted a comparative analysis of computational complexity across different model dimensions and software implementations. Specifically, we assessed how factors including layer depth, neuron width, activation functions, and framework choice impact processing time. We contrasted the DCNN, which incorporates analytical derivative calculations, against a standard MLP baseline without derivative consideration.

\begin{landscape}
\thispagestyle{empty}
\begin{table}[htbp]
\begin{center}
\begin{minipage}{185mm}
\caption{Comparison of computational time by the configurations of neural networks and packages, displayed as mean values with standard deviation in seconds. The other settings is similar to that in Section \ref{subsection: Experimental design}$^{\rm a}$.}
{\begin{tabular}{@{}ccccrlrlrlrl}
    \hline
    \multicolumn{4}{c}{Network configuration (\# of)} & \multicolumn{2}{c}{\multirow{2}{*}{MLP(Pytorch)}} & \multicolumn{2}{c}{\multirow{2}{*}{DCNN(Pytorch)}} & \multicolumn{2}{c}{\multirow{2}{*}{MLP(JAX)}}& \multicolumn{2}{c}{\multirow{2}{*}{DCNN(JAX)}}\\
    \multirow{2}{*}{\begin{tabular}{c}Layers\\$(L)$\end{tabular}}&\multirow{2}{*}{\begin{tabular}{c}Neurons\\$(d_l)$\end{tabular}}&\multirow{2}{*}{\begin{tabular}{c} Act. func.\\$(f)$\end{tabular}}& \multirow{2}{*}{\begin{tabular}{c}Param.\\$(W, b)$\end{tabular}}&&&&&&&& \\
    &&&& mean & (std) & mean & (std) & mean & (std) & mean & (std) \\
    \hline
    3 & 16 & softplus & 337 & 35.21 & (0.34) & 69.57 & (0.67) & 9.23 & (0.45) & 11.75 & (0.59) \\
    3 & 16 & tanh & 337 & 29.76 & (0.59) & 63.12 & (0.37) & 9.54 & (1.74) & 12.16 & (0.53) \\
    3 & 16 & elu & 337 & 33.35 & (0.12) & 68.90 & (1.61) & 9.79 & (1.69) & 12.40 & (0.46) \\
    \cdashline{2-12}3 & 64 & softplus & 4,417 & 62.10 & (0.18) & 152.36 & (0.55) & 9.32 & (0.69) & 11.97 & (0.48) \\
    3 & 64 & tanh & 4,417 & 47.85 & (0.48) & 137.84 & (1.97) & 9.52 & (1.45) & 12.00 & (0.39) \\
    3 & 64 & elu & 4,417 & 59.47 & (2.41) & 159.02 & (8.75) & 9.74 & (1.47) & 12.14 & (0.68) \\
    \cdashline{2-12}3 & 128 & softplus & 17,025 & 102.56 & (0.78) & 368.57 & (5.96) & 9.15 & (0.61) & 12.14 & (0.54) \\
    3 & 128 & tanh & 17,025 & 85.68 & (0.96) & 341.05 & (5.38) & 9.43 & (1.29) & 11.94 & (0.62) \\
    3 & 128 & elu & 17,025 & 98.83 & (0.52) & 372.14 & (2.94) & 9.62 & (1.48) & 12.14 & (0.60) \\
    \cdashline{1-12}5 & 16 & softplus & 881 & 68.13 & (0.19) & 148.75 & (0.38) & 11.76 & (0.57) & 17.90 & (0.61) \\
    5 & 16 & tanh & 881 & 56.19 & (0.21) & 135.05 & (0.29) & 12.05 & (0.38) & 17.15 & (0.40) \\
    5 & 16 & elu & 881 & 63.75 & (0.30) & 143.70 & (0.56) & 12.17 & (1.44) & 18.02 & (0.47) \\
    \cdashline{2-12}5 & 64 & softplus & 12,737 & 133.15 & (0.78) & 381.98 & (2.24) & 12.57 & (1.54) & 16.97 & (0.42) \\
    5 & 64 & tanh & 12,737 & 102.64 & (0.30) & 344.19 & (2.87) & 12.29 & (1.46) & 17.06 & (0.48) \\
    5 & 64 & elu & 12,737 & 125.98 & (1.60) & 385.76 & (2.51) & 12.42 & (1.38) & 17.28 & (0.40) \\
    \cdashline{2-12}5 & 128 & softplus & 50,049 & 245.30 & (2.50) & 1,021.43 & (7.08) & 12.44 & (1.39) & 17.51 & (0.51) \\
    5 & 128 & tanh & 50,049 & 212.71 & (3.85) & 948.21 & (9.75) & 11.80 & (1.58) & 17.62 & (0.50) \\
    5 & 128 & elu & 50,049 & 248.08 & (4.03) & 1,020.47 & (5.92) & 12.27 & (0.55) & 17.56 & (0.49) \\
    \cdashline{1-12}9 & 16 & softplus & 1,969 & 133.77 & (0.26) & 305.85 & (0.30) & 16.36 & (0.42) & 28.26 & (0.59) \\
    9 & 16 & tanh & 1,969 & 109.51 & (0.21) & 278.17 & (0.35) & 16.51 & (1.62) & 27.99 & (0.60) \\
    9 & 16 & elu & 1,969 & 124.08 & (0.50) & 294.36 & (0.36) & 15.17 & (0.47) & 25.28 & (0.83) \\
    \cdashline{2-12}9 & 64 & softplus & 29,377 & 287.31 & (12.13) & 897.94 & (41.08) & 17.06 & (0.45) & 26.80 & (0.87) \\
    9 & 64 & tanh & 29,377 & 212.87 & (1.70) & 762.05 & (5.53) & 17.14 & (1.42) & 27.22 & (0.93) \\
    9 & 64 & elu & 29,377 & 269.09 & (8.48) & 871.80 & (26.58) & 15.91 & (0.87) & 26.67 & (0.91) \\
    \cdashline{2-12}9 & 128 & softplus & 116,097 & 541.15 & (16.32) & 2,252.24 & (28.87) & 17.50 & (1.52) & 29.21 & (0.59) \\
    9 & 128 & tanh & 116,097 & 485.25 & (24.02) & 2,260.50 & (106.55) & 16.72 & (1.77) & 28.30 & (0.45) \\
    9 & 128 & elu & 116,097 & 554.98 & (30.61) & 2,436.35 & (135.17) & 15.88 & (0.61) & 28.22 & (0.84) \\
    \hline
\end{tabular}}
\footnotesize{$^{\rm a}$ Experimental conditons are $N=575, M=286$ and $10,000$ epochs repeated 10 times in the synthesized data with $\rho=-0.4$ and $\nu=0.6$ in the SABR model. \\ $^{\rm b}$ In the case of the ELU function, $\alpha = \beta = 1$ as applied in Table \ref{table:activation_function_derivatives}.
}
\label{table: computational time}
\end{minipage}
\end{center}
\end{table}
\vfill
\raisebox{-10pt}{\makebox[\linewidth]{\thepage}}
\end{landscape}

The results in Table \ref{table: computational time} demonstrate substantially stable overhead for the DCNN versus MLP, with the incorporation of derivative information consistently increasing computation by approximately 3-4 times. This aligned with the additional mathematical operations required for derivative computations per Eqs. \eqref{eq: cost function} and \eqref{eq: definition of nabla and nabla squeared}. Meanwhile, standard neural network stacking entailed exponential complexity growth with expanding width and depth.

The comparative assessment also revealed pronounced performance advantages conferred by the JAX framework leveraging just-in-time compilation and GPU acceleration over PyTorch. Across most configurations, JAX reduced execution time by three to four times. For larger models, speedups reached up to 20 times, likely due to superior numerical stability from optimized code generation targeting parallel hardware. Diverging convergence behaviour on expansive models also suggests JAX may confer superior numerical stability. The consistent Derivative-Constrained Neural Network profile with model size makes this architecture appealing for extensible large-scale machine learning applications compared to standard multilayer perceptron performance degradation.

Based on the analysis, mid-sized DCNN models with 2-4 layers and 16-64 units balance predictive accuracy and feasible computation in under five seconds. Larger models should leverage JAX for optimal efficiency and stability. Overall, DCNN presents a realistic tool for practitioners through automatic differentiation, enabling controlled derivative computation alongside the utilization of a recent, efficient programming package.

\section{Backtests of empirical data from 1 October 2022 to 30 September 2023}
\label{appendix: Backtests}
This section conducts robust backtesting on extensive intraday options data for DCNN with the dataset on S\&P 500 options spanning the period from 2022 to 2023. To investigate the efficiency of the DCNN for the historical data, the backtests were conducted using intraday traded prices of S\&P 500 options for 248 days from 1 October 2022 to 30 September 2023; we obtained about $500,000$ points on a daily basis via CBOE DataShop\footnote{CBOE DataShop. (2023). SPX options. https://datashop.cboe.com/option-trades. [dataset]} as summarized in Table \ref{table: SP500 options cboe}. We added the synthesized points corresponding to boundary conditions to training (note not for a statistics analysis), and all other setups are the same as in \ref{subsection: Experimental design}.

\begin{table}[htbp]
\begin{center}
\caption{A statistics of intraday prices of S\&P 500 options for 248 business days from 1 October 2022 to 30 September 2023.}
{\small\begin{tabular}{@{}lrlrrl}
    \toprule
        {\small[per day]} & \small{Mean} & \scriptsize{(std. dev.)} & \small{Median} & \scriptsize{(min.)} & \scriptsize{(max.)} \\
    \midrule
        all count of intraday traded price & 527,646 & \scriptsize{(71,116)} & 528,771 & \scriptsize{(183,418)} & \scriptsize{(734,441)} \\
        \# of unique grids & 6,516 & \scriptsize{(530)} & 6,458 & \scriptsize{(4,580)} & \scriptsize{(8,269)} \\
        \% of call options & 40.20 \%  & \scriptsize{(2.02 \%)} & 40.01 \% & \scriptsize{(35.16 \%)} & \scriptsize{(47.09 \%)} \\
        \% of short term ($\tau < 1M$) & 95.47 \% & \scriptsize{(0.93 \%)} & 95.61 \% & \scriptsize{(91.40 \%)} & \scriptsize{(97.07 \%)} \\
        \% of long term ($\tau > 1Y$) & 2.69 \% & \scriptsize{(0.52 \%)} & 2.62 \% & \scriptsize{(1.42 \%)} & \scriptsize{(4.32 \%)} \\
        \% of near ATM ($M \in [0.9, 1.1]$) & 71.97 \% & \scriptsize{(2.66 \%)} & 72.10 \% & \scriptsize{(64.43 \%)} & \scriptsize{(76.95 \%)} \\
        \% of far OTM ($M \notin [0.5, 1.5]$) & 2.00 \% & \scriptsize{(0.33 \%)} & 1.99 \% & \scriptsize{(1.17 \%)} & \scriptsize{(3.05 \%)} \\
    \bottomrule
\end{tabular}}
\label{table: SP500 options cboe}
\end{center}
\end{table}
Analysis of the data presented in Table \ref{table: SP500 options cboe} reveals that intraday traded data exhibits an uneven distribution in the input variables. Approximately 95\% of trades occur in short-dated options (within 1 month), while over 70\% are concentrated around at-the-money (ATM) strikes. This introduces a relatively challenging calibration problem compared to more general cases. Effective modelling needs flexible interpolation ability for sparsely populated points alongside the incorporation of additional mesh information encoding derivative constraints within the DCNN framework.

\begin{figure}[htbp]
    \centering
    \includegraphics[width=1.0\columnwidth]{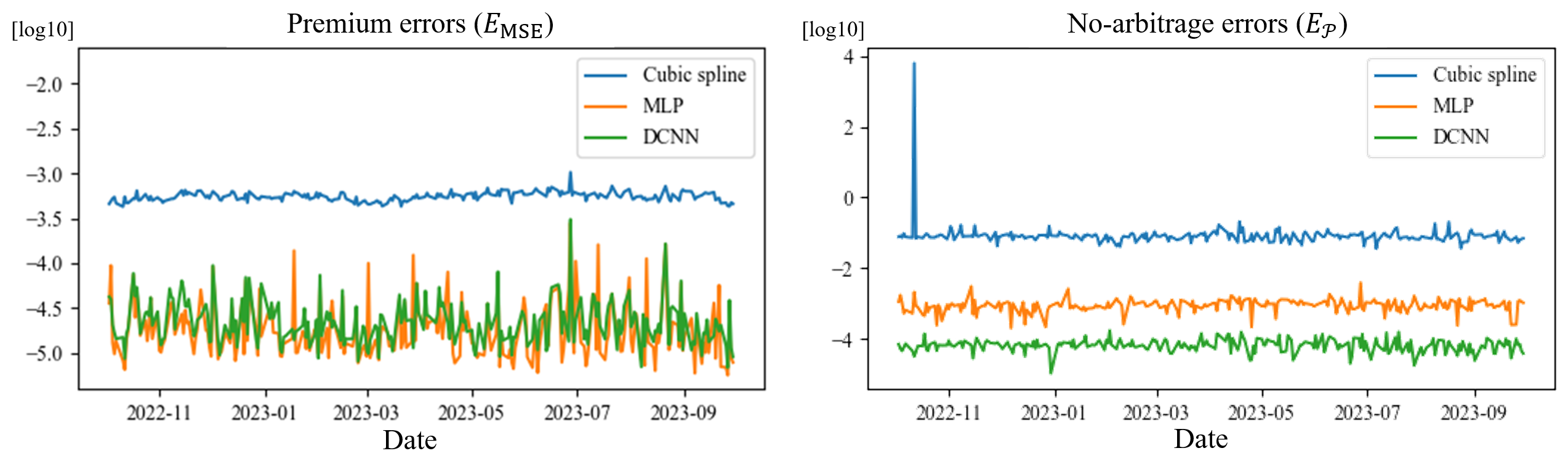}
    \caption{Backtests of intraday traded prices of S\&P 500 options for 248 days from 1 October 2022 to 30 September2023.}
    \label{Figure: NoArbSS_cboe_backtests}
\end{figure}

Results in Figure \ref{Figure: NoArbSS_cboe_backtests} demonstrate the flexibility of DCNNs to accurately fit both the cross-section of market prices as well as fluctuations over time. The requirements for an efficient calibration are reconciling the differing distributions implied under risk-neutral and real-world measures while capturing higher moments like skewness and kurtosis.

When there is a model that has explanatory power for both probability measures, it is possible to obtain a more accurate probability distribution using historical data, and it will also be possible to perform derivative evaluation and risk management in a consistent manner. However, in the formulation based on historical data, there is a degree of freedom in determining the so-called market price of risk. In the empirical backtests in this section, we observed the DCNNs show stability and efficiency in the errors of whole periods compared with other models. The results support that DCNNs surgically incorporate derivative information in the loss function to enhance pricing and risk estimations without historical statistical assumptions, or parameters of stochastic models.

\section{Self-Adaptive weighting for Derivative-Constrained Neural Network}
\label{appendix: Self-Adaptive Derivative-Constrained Neural Network}
DCNNs with weighted loss terms have improved stability and accuracy over MLPs, yet remain rigid and non-adaptive to the weighting terms. In alignment with the neural network philosophy of self-adaptation, this section proposes a straightforward procedure applying fully trainable weights to generate multiplicative soft-weighting and attention mechanisms, following in \cite{mcclenny2020self}. Rather than hard-coding weights at specific parts of the loss, this proposed self-adaptive DCNN updates the loss function weights via gradient descent concurrently with the network weights. Self-adaptive DCNN employs the following weighting function in loss function based on Eq \eqref{eq: cost function}:
\begin{equation}
\lambda (m, \mathbf{x}) = \begin{cases} \gamma(m) \cdot g(\mathbf{x}), & \text{if penalty} \\ 0, & \text{if not penalty}\end{cases},
\end{equation}
where $m \in \mathbb{R}$ are still constants and $\gamma(x)$ are intensifier functions, which is monotonically increasing. The objective is minimized the total cost with respect to the network weights and bias but also is maximized with respect to the self-adaptation weights $m$. Considering the updates of a gradient ascent approach of learning with $k$-th weight vector of $i$-th derivative $m_i(k)$, $k=1, \ldots, I_{max}$, 
\begin{equation}
\begin{split}
m_i^{(j)}(k+1) = m_i^{(j)}(k) + \eta_{m_i^{(j)}} \nabla_{m_i^{(j)}} E(\mathbf{X}, & \hat{\mathbf{X}}, \Phi),
\end{split}
\end{equation}
where $\eta > 0$ is the learning rate for self-adaption weights. In the learning step, the derivatives with respect to self-adaption weights are increased when it breaks the constraints of derivative terms respectively,
\begin{equation}
\begin{split}
\nabla_{m_i^{(j)}} E(\mathbf{X}, & \hat{\mathbf{X}}, \Phi) = \begin{cases} \gamma'(m_i^{(j)}) \left\{ g( h_1 \nabla\Phi(\hat{\mathbf{x}}^{(j)})) + g( h_2 {\nabla}^{2}\Phi(\hat{\mathbf{x}}^{(j)})) \right\}, & \text{if penalty} \\ 0, & \text{if not penalty}\end{cases}.
\end{split}
\end{equation}
Rather than being selected a priori, the self-adaptation weights and mask values produce penalty costs that increase adaptively. These penalty costs are not prescribed beforehand but are updated dynamically through the neural network training procedure.

\begin{figure}[H]
    \centering
    \includegraphics[width=0.8\columnwidth]{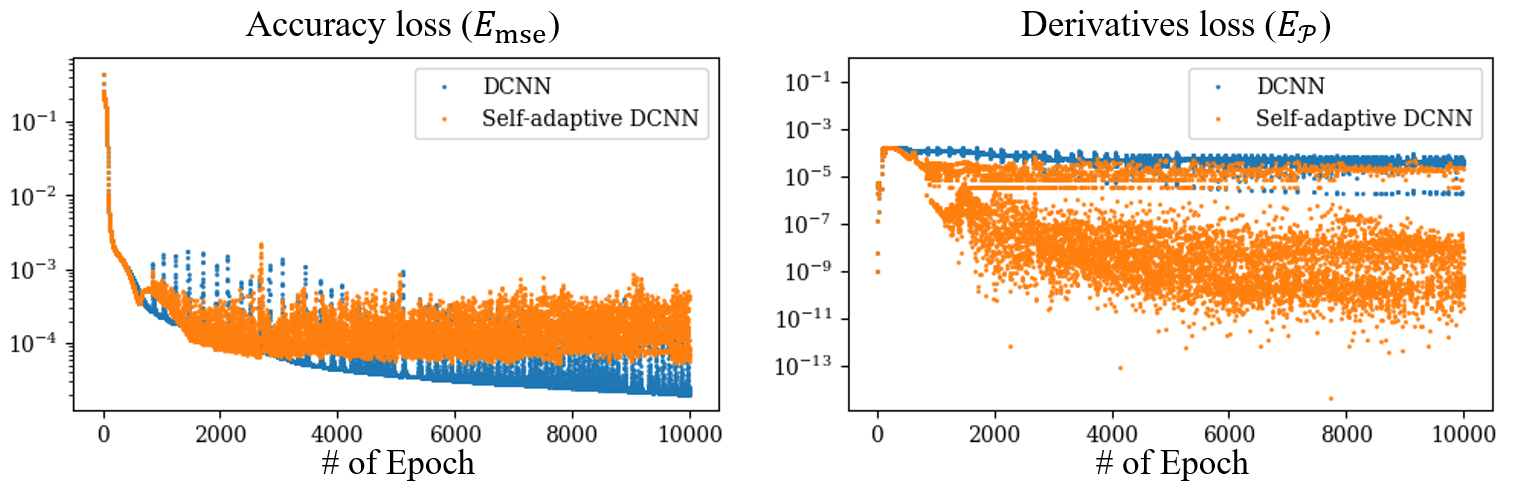}
    \caption{The accuracy ($E_\text{MSE}$) and derivative loss ($E_p$) in self-adaptive DCNN with a logarithmic scale by learning epochs.}
    \label{Figure: NoArbSS_learning_results_sa}
\end{figure}

As a testing result in this section, we applied $x^2$ as an intensifier function $\gamma(x)$, $m_1, m_3=0.001$, and $m_2=0.01$.
Other experimental setups are similar to that in Section \ref{subsection: Experimental design} {\small($N=575, M=286$ and $10,000$ epochs, in the synthesized data with $\rho=-0.4$ and $\nu=0.6$ in the SABR model)}.
We also observed weight distributions after training in Figure \ref{Figure: NoArbSS_learning_results_sa}.

\begin{figure}[htbp]
    \centering
    \includegraphics[width=1.0\columnwidth]{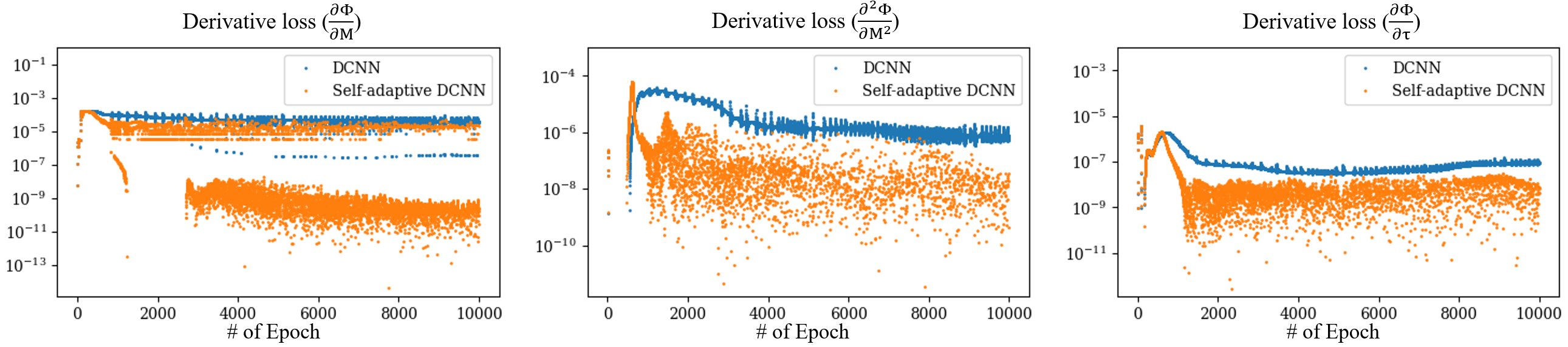}
    \caption{The derivatives losses that are elements to the derivatives ($E_p$) in Figure \ref{Figure: NoArbSS_learning_results_sa} with a logarithmic scale by learning epochs.}
    \label{Figure: NoArbSS_learning_results_sa_derivs}
\end{figure}

The proposed self-adaptive DCNN architecture was implemented and evaluated using the previously described datasets. Performance metrics over training epochs reveal smooth weight convergence without instability. Learning dynamics are visualized in Figures \ref{Figure: NoArbSS_learning_results_sa} and \ref{Figure: NoArbSS_learning_results_sa_derivs}. Results demonstrate the soft-weighting mechanism successfully reduces derivative losses, although some trade-off with accuracy loss is exhibited. Additionally, analysis shows that all derivative components are equally influenced by the soft-weighting rather than only subsets of the elements.

\begin{figure}[htbp]
    \centering
    \includegraphics[width=1.0\columnwidth]{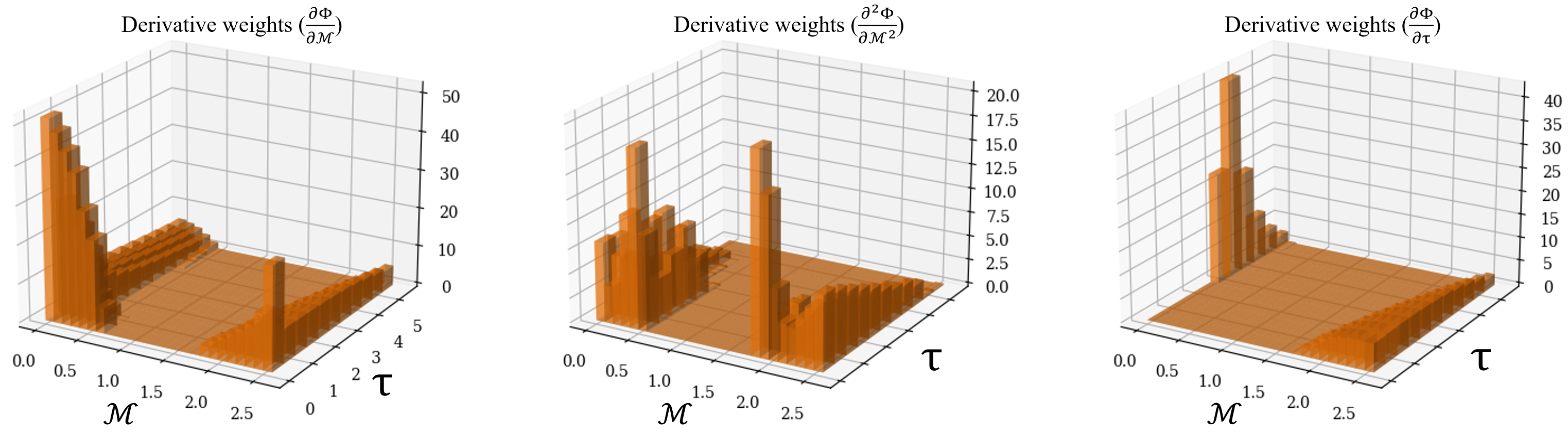}
    \caption{The distributions of weights for each derivative loss after learning in self-adaptive DCNN.}
    \label{Figure: NoArbSS_learning_results_sa_derivs_weights}
\end{figure}

Figure \ref{Figure: NoArbSS_learning_results_sa_derivs_weights} illustrates the distributions of loss weights for each derivative after training in the self-adaptive DCNN. High weight values indicate areas where no-arbitrage conditions were more likely to be violated during learning. Analysis reveals clustered high-loss weighting near boundary conditions and low-value short-dated options.

These results demonstrate the benefits of incorporating dynamic, trainable loss weights. By enabling the network to automatically learn optimal weightings alongside normal connection weights, adherence to derivative constraints improves without requiring pre-determined weight configurations. The performance gains validate that allowing neural networks to self-direct appropriate data-dependent weighting strategies outperforms manual loss function engineering. This aligns with core neural network principles. Just as base connections adapt during training, letting loss contribution weights calibrate automatically is advantageous. Making the weight variables differentiable parameters learned via backpropagation allows the network to holistically optimize all components in unison. This confers greater flexibility to find more accurate solutions with fewer parameters.

\end{document}